\documentclass[pre,showpacs,twocolumn,amsmath,amssymb]{revtex4}
\usepackage{graphicx}% Include figure files
\usepackage{dcolumn}% Align table columns on decimal point
\usepackage{bm}% bold math

\begin{document}

%\preprint{APS/123-QED}

\title{Back-stepping, hidden substeps, and conditional dwell times in molecular motors}

\author{Denis Tsygankov}
 \email{dtsygank@umd.edu}
\author{Martin Lind\'en}
 \email{linden@kth.se}
 \altaffiliation[Permanent address: ]{Department of Theoretical Physics, Royal Institute of Technology, AlbaNova, 106 91 Stockholm, Sweden}
\author{Michael E. Fisher}
 \email[Corresponding author: ]{xpectnil@umd.edu}
 \affiliation{Institute for Physical Science and Technology, University of Maryland, College Park, Maryland, 20742, USA}

\date{\today}

\begin{abstract}
Processive molecular motors take more-or-less uniformly sized steps,
along spatially periodic tracks, mostly forwards but increasingly
backwards under loads. Experimentally, the major steps can be
resolved clearly within the noise but one knows biochemically that
one or more mechanochemical substeps remain hidden in each enzymatic
cycle. In order to properly interpret experimental data for
back/forward step ratios, mean conditional step-to-step dwell times,
etc., a first-passage analysis has been developed that takes account
of hidden substeps in $N$-state sequential models. The explicit,
general results differ significantly from previous treatments that
identify the observed steps with complete mechanochemical cycles;
e.g., the mean dwell times $\tau_+$ and $\tau_-$ prior to forward
and back steps, respectively, are normally {\it unequal} although
the dwell times $\tau_{++}$ and $\tau_{--}$ between {\it successive}
forward and back steps are equal. Illustrative ($N=2$)-state
examples display a wide range of behavior. The formulation extends
to the case of two or more detectable transitions in a multistate
cycle with hidden substeps.
\end{abstract}


%\keywords{Suggested keywords}%Use showkeys class option if keyword
                              %display desired
\maketitle

%***********************************************************************************************************************************
\section{\label{sec:level1}Introduction}

Processive motor proteins or molecular motors~\cite{howard,bray}
(such as kinesin, cytoplasmic dynein, and myosin V) ``walk'' along
molecular tracks (microtubules and actin filaments) taking {\it
observed mechanical steps} of well defined (mean) spacing $d$ , each
step being  of {\it ``negligibly short''} ($\lesssim 100\,\mu s$)
duration relative to the mean time(s) between steps that are of
order $1$ to $20\,\textrm{ms}$
~\cite{howard,bray,mehta99,rief00,nishiyama02,block03,nishiyama03,yildiz03,snyder04,uemura04,baker04,oiwa05,carter05,taniguchi05,clemen05,guydosh06,toba06}.
Steps may be taken {\it forwards} ($+$) or {\it backwards} ($-$).
The mean velocity (observed over 10's to 100's ) of steps, $V$, is a
{\it function} of the load $\textbf{F}=(F_x,F_y,F_z)$ exerted on the
motor and the fuel concentration [ATP] (for most
cases)~\cite{fisher99,kolomeisky03,fisher05} and, in general, of
other features of the aqueous solution including the pH, ionic
strength, temperature $T$, and other reagents/reactants such as
[ADP], [Pi], [AMP-PNP], [BeF$_2$],
etc.)\cite{howard,block03,carter05,guydosh06}.

Motor proteins  are enzymatic {\it catalysts} that, following
biochemical knowledge and principles, turn over one ``fuel molecule"
(usually ATP) for each full step via (in the simplest case) a {\it
linear sequence of reversible kinetic transitions} (or {\it
reactions}) embodying $N$ (bio)chemical states per
turnover~\cite{fisher99,kolomeisky03,fisher05,fisher99a,kolomeisky00,kim05,bustamante01}.
This situation is embodied in the following {\it basic sequential
model}
\begin{align}
&\cdots\begin{array}{c}
    u_{N-1}\\
    \xleftarrow{\quad\quad\quad}\\
    \vspace{-10 mm}\\
    \xrightarrow{\quad\quad\quad}\\
    w_{0\equiv N}
\end{array}
[0]_l\begin{array}{c}
    u_0\\
    \xleftarrow{\quad\quad\quad}\\
    \vspace{-10 mm}\\
    \xrightarrow{\quad\quad\quad}\\
    w_1
\end{array}
(1)_l\begin{array}{c}
    u_1\\
    \xleftarrow{\quad\quad\quad}\\
    \vspace{-10 mm}\\
    \xrightarrow{\quad\quad\quad}\\
    w_2
\end{array}
(2)_l\begin{array}{c}
    u_2\\
    \xleftarrow{\quad\quad\quad}\\
    \vspace{-10 mm}\\
    \xrightarrow{\quad\quad\quad}\\
    w_3
\end{array}
\cdots\nonumber\\
&\cdots\begin{array}{c}
    u_{N-2}\\
    \xleftarrow{\quad\quad\quad}\\
    \vspace{-10 mm}\\
    \xrightarrow{\quad\quad\quad}\\
    w_{N-1}
\end{array}
(N-1)_l\begin{array}{c}
    u_{N-1}\\
    \xleftarrow{\quad\quad\quad}\\
    \vspace{-10 mm}\\
    \xrightarrow{\quad\quad\quad}\\
    w_N
\end{array}
[N]_l\equiv[0]_{l+1}\begin{array}{c}
    u_0\\
    \xleftarrow{\quad\quad}\\
    \vspace{-10 mm}\\
    \xrightarrow{\quad\quad}\\
    w_1
\end{array}
\cdots,\label{BasicModel}
\end{align}

%\begin{equation}\label{BasicModel}
%    \begin{array}{l}
%        \cdots\autorightleftharpoons{$u_{N-1}$}{$w_{0\equiv N}$}[0]_l\autorightleftharpoons{$u_0$}{$w_1$}(1)_l\autorightleftharpoons{$u_1$}{$w_2$}(2)_l\autorightleftharpoons{$u_2$}{$w_3$}\cdots\\
%        \vspace{-5 mm}\\
%        \cdots\autorightleftharpoons{$u_{N-2}$}{$w_{N-1}$}(N-1)_l\autorightleftharpoons{$u_{N-1}$}{$w_N$}[N]_l\equiv[0]_{l+1}\autorightleftharpoons{$u_0$}{$w_1$}\cdots,
%    \end{array}
%\end{equation}
which is understood to repeat periodically as the motor moves
processively along its track. The subscript $l=1,2,\cdots$ labeling
the $N$ basic states $[0]$, $(1)$, $\cdots(N-1)$, denotes the sites
on the linear track spaced at distance $d$ apart.

By convention the state $(i)=[0]$ is ``bound'' or
``nucleotide-free''  so that the transition $[0]\rightarrow(1)$
represents the {\it binding} of one fuel molecule to the awaiting
motor. Thus we write
\begin{equation}
u_0=k_0[\mbox{ATP}],\label{ATPdepend}
\end{equation}
where the pseudo-first-order rate constant $k_0$ and all the
remaining rate constants $u_i$, $w_j$ depend also on $\textbf{F}$.
But under fixed conditions ($\textbf{F}$, [ATP], $\cdots$), the
rates do not change. Note that this formulation embodies the {\it
tight coupling principle} of one fuel molecule being consumed per
(forward) step~\cite{howard,fisher99,kolomeisky03,bustamante01}.
This is {\it assumed} in the basic model, which also neglects
irreversible {\it detachments} from the track (which, however, can
be included readily in
principle~\cite{fisher99,fisher99a,kolomeisky00}).

When convenient we will allow the state labels $i$, $j$, $\cdots$ to
take values outside the basic range $[0,N-1]$; for that reason we
adopt the periodicity convention
\begin{equation}
u_{i+lN}\equiv u_i,\;\;w_{j+lN}\equiv
w_j,\;\;l=0,\pm1,\pm2,\cdots.\label{Convention}
\end{equation}

Now, in the simplest experimental situation, as observed for
kinesin, no {\it mechanical substeps} are
detected~\cite{nishiyama02,carter05} to within the noise level
(which amounts to $\Delta x\lesssim 1\,\textrm{nm}$). Furthermore to
within the resolution time ($\lesssim 100\,\mu s$), successive steps
occur at times, say, $\cdots, t_{k-1}, \;t_k, \;t_{k+1}, \cdots$.
Thus, between the identifiable mechanical steps of (mean) magnitude
$d$, the motor {\it dwells} in a {\it mechanical state} that, within
the noise level $\Delta x$, appears well defined with no
systematically detectable {\it substeps}, forwards or backwards.
Then, individual {\it dwell times} in the mechanical states, namely,
\begin{equation}
\tau_k=t_k-t_{k-1},\label{DefDwellTimes}
\end{equation}
can be measured to reasonable precision and averages may be
computed, over ``many'' observations encompassing, say, $n$ steps,
to yield an {\it overall mean dwell time}
\begin{equation}
\tau=\left<\tau_k\right>\approx\frac{1}{n}\sum_{k=1}^n\tau_k.\label{OverallMeanDwellTimes}
\end{equation}
Here and below we use the ``asymptotically equals'' symbol $\approx$
to indicate an approximate equality that becomes exact in a long run
under steady-state conditions.

Given a (sufficiently long) sequence of $n$ observed steps with
$n_+$ forward steps and $n_-$ backward steps, we can also define the
(steady-state) {\it step splitting probabilities} or back-step and
forward-step fractions
\begin{equation}
\pi_+\approx n_+/n,\qquad\pi_-\approx n_-/n,\label{StepSplProb}
\end{equation}
where, since $n=n_++n_-$, one has
\begin{equation}
\pi_++\pi_-=1.\label{SumOfProbIsOne}
\end{equation}

Furthermore, dwell times before a $+$ or $-$ step can be (and have
been~\cite{carter05}) measured separately leading to distinct {\it
prior dwell times}
\begin{equation}
\tau_+=\frac{1}{n_+}\mbox{$\sum^{+}$}\tau_k\qquad\mbox{and}\qquad\tau_-=\frac{1}{n_-}\mbox{$\sum^{-}$}\tau_k,\label{PriorDwellTimes}
\end{equation}
the restricted sums including just $+$ or $-$ steps, respectively.

To the degree that the runs are long so that $\pi_+$ and $\pi_-$ may
be accurately considered as probabilities one must evidently also
have
\begin{equation}
\pi_+\tau_++\pi_-\tau_-=\tau.\label{TotalDwellasSum}
\end{equation}

As discussed recently in some detail~\cite{fisher05,kim05}, each
individual biochemical state $(i)$, may be characterized by a
definite (mean) {\it longitudinal location} in physical space, i.e.,
{\it along} the track, which we supposed aligned with the $x$
coordinate, and, possibly, {\it transverse} to the track, the
$y$-coordinate, or {\it normal} to the track, the $z$-coordinate.
Hence the basic model implies the existence of {\it substeps}, say,
of magnitude
\begin{equation}
d_j=x_{j+1}-x_j,\label{StepMagnitude}
\end{equation}
between successive mechanochemical states~\cite{foot0}. However, the
great majority of these mechanical displacements will be {\it
hidden} by noise and so {\it unobservable}. This is the crucial
issue.

The evidence (in particular for kinesin~\cite{nishiyama02,carter05})
reveals the existence of one principal or {\bf major mechanical
substep} of magnitude
\begin{equation}
d_M=x_{M+1}-x_M\simeq d,\label{MajorMechStep}
\end{equation}
that corresponds to a specific transition
$(M)\!\!\rightarrow\!\!(M+1)$ for a {\it forward} or $+$ {\it step}.
Such a unique forward-step is sometimes called a ``power stroke''.
Then, clearly, within the basic model a back-step ($-$) corresponds
to  the specific transition $(M+1)\!\rightarrow\!(M)$.

For simplicity we will initially suppose that there is {\it only
one} such single, well defined and observable principal mechanical
transition in the processive reaction cycle: it will be referred to
as a {\bf major transition} while all other smaller, unobservable
displacements, presumed ``hidden,'' will be termed {\it substeps}.

It is of interest, all the same, to analyze situations in which,
within the full cycle, there are a number of {\it visible} (or
observable) {\it substeps}. Indeed, an initial substep large enough
to be readily observable was predicted for myosin V by Kolomeisky
and Fisher~\cite{kolomeisky03} on the basis of dwell-time data
obtained at different [ATP] and force levels~\cite{mehta99}; it was
later observed unambiguously by Uemura et al~\cite{uemura04}. Thus,
in Sec.~\ref{sec:level5} below, the case of $K$ ($<N$) distinct
major substeps is considered explicitly~\cite{foot0A}.

Nevertheless, since most of the forward and reverse transitions,
$(i)\rightarrow(i-1)$ and $(i-1)\rightarrow(i)$, are {\it not}
observable, one does {\it not} know (and {\it cannot} tell) the
(bio)chemical (sub)state of the motor {\it during an observed dwell
time}, $\tau_k$, between steps $k-1$ and $k$:  see
(\ref{DefDwellTimes}).  Indeed, the biochemical state will change as
time progresses and not necessarily in a uniform sense, e.g., ATP
might bind and then be released (or unbind) without undergoing a
hydrolysis step. For this {\it crucial reason} once the basic model
has $N=2$ or more states the expressions for $\tau_+$ and $\tau_-$
in terms of the basic rates $u_i$ and $w_j$ {\it cannot be trivial}
--- and the same goes for the splitting or backwards and forwards probabilities
$\pi_+$ and $\pi_-$.

The basic theoretical problem is thus to find {\it explicit
expressions} for $\pi_+$, $\pi_-$, and for the {\it partial dwell
times} as well as for {\it conditional} or {\it pairwise} stepping
fractions, $\pi_{++}$, $\pi_{-+}$, $\cdots$, and dwell times,
$\tau_{++}$, $\tau_{-+}$, $\cdots$, that it is natural to introduce
(as seen below). Indeed, it is clear that these general statistical
concepts are not restricted to linear or translocational motors, on
which we have focussed; in fact, they apply equally to {\it rotary
stepping motors} like
F$_1$-ATPase~\cite{yasuda98,shimabukuro03,nishizaka04,shimabukuro05},
F$_1$F$_0$-ATPase~\cite{diez04,ueno05}, and bacterial flagellar
motors~\cite{bray,sowa05}. However, since, in these respects,
kinesin and myosin V have been studied more extensively, we will
retain the language appropriate for processive motor proteins
walking on linear tracks.

In previous theoretical
studies~\cite{kolomeisky03,kolomeisky05,qian06,wang06} splitting
probabilities and conditional mean dwell times have been introduced
in the context of molecular motors {\it via} the following
definitions (which are distinguished from those used in the
discussion above by haceks) specifically:
$\check{\pi}_+\equiv\pi_{0,N}$ and $\check{\pi}_-\equiv\pi_{0,-N}$
are the probabilities that a motor starting at a well defined
physical site $l$ along the track in the binding state $(j)=[0]$
will arrive at the next site $l+1$ or back at the previous site
$l-1$ (in both cases in state $(j)=[0]$) without having undergone
the opposite transition to complete a full backward or forward
cycle, respectively. Then, similarly,
$\check{\tau}_{\pm}\equiv\tau_{0,\pm N}$ are the average times a
motor spends at site $l$ (starting at $j=0$) before completing a
full forward or backward cycle to site $l\pm 1$ (in state
$(j)=[0]$). Exact results for such statistics can be derived by
mapping into a Markov renewal process~\cite{qian06,wang06}.

With these definitions the explicit formulae
obtained~\cite{kolomeisky03,kolomeisky05} for
$\;\check{\pi}_{\pm}\;$ and $\;\check{\tau}_{\pm}\;$ correspond to a
{\it full mechanochemical cycle} during which a complete forward or
backward step is certainly taken. However, the resulting expressions
can be applied to the analysis of experimental stepping data {\it
only if} these data allow one to identify each full cycle. If,
instead, the observed noise hides one or more of the $N>1$
biochemical or mechanochemical substeps, while only major mechanical
transitions are detectible, one {\it cannot} in general decide
unambiguously whether a motor executed the detectable steps (or
power strokes) with or without completing a full cycle. In such
cases, the previous expressions cannot be applied to account for the
observed step fractions and dwell times $\pi_{\pm}$ and
$\tau_{\pm}$. Instead, the results must be modified to allow for the
ambiguity arising from the hidden substeps. It transpires, as we
show below, that this rather subtle and at first-sight
inconsequential small difference actually leads to significant
changes in the load dependence especially (but not exclusively) when
the fractions of back-steps and forward-steps are similar in
frequency, i.e., on approaching stall conditions when the velocity,
$V$, becomes small relative to its load-free
value~\cite{fisher99,kolomeisky03,fisher05}.

\begin{figure*}
\includegraphics[angle=0, scale=0.70]{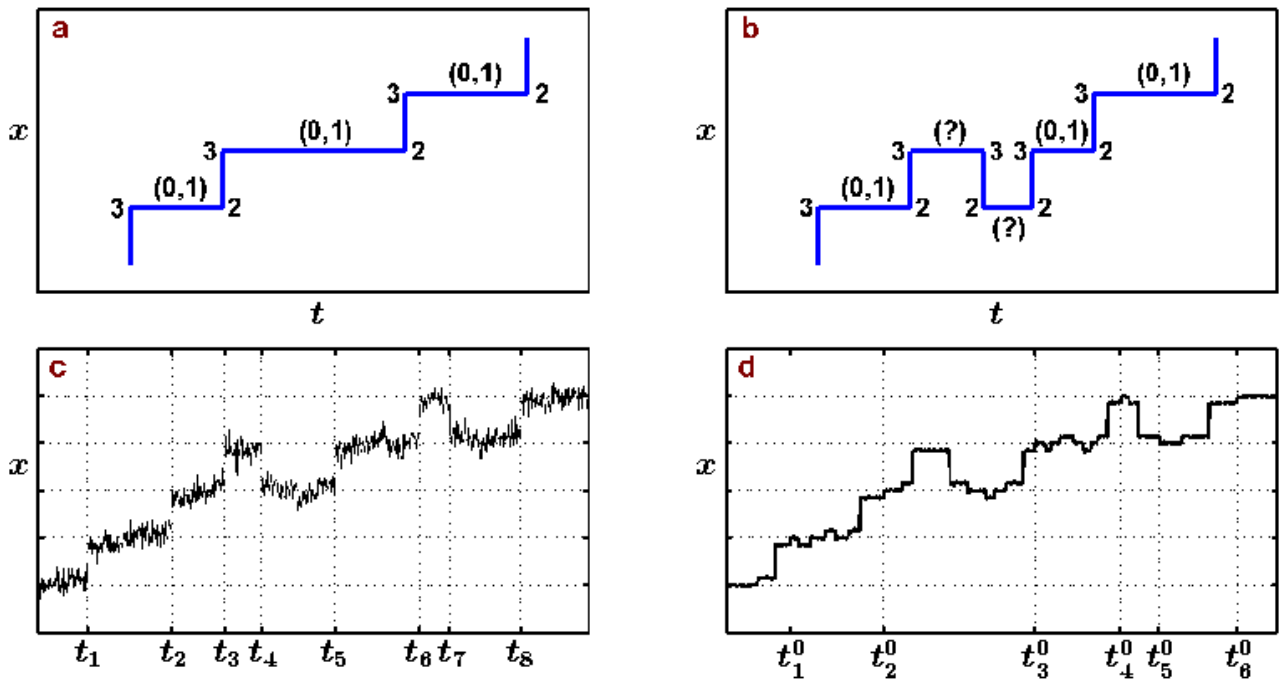}
\caption{\label{fig:wide} Schematic graphs illustrating why the
full-cycle interpretation is not adequate for describing stepping
data: Plot {\bf (a)} depicts a hypothetical time series of forward
steps for an ($N=4$) cycle with $M=2$. The (0,1)'s indicate that the
states $[0]_l$ and $(1)_l$ are both certain to occur at least once
between the pairs of states $(3)_{l-1}$ and $(2)_l$. (See the text.)
Then {\bf (b)} represents a similar time series but with one major
back-step. The question-marks between the forward-back and
back-forward steps indicate that one does not know if the motor has
ever passed through the states $[0]_l$ and/or $(1)_l$ in these
intervals. Finally, {\bf (c)} and {\bf (d)} depict two equivalent
time series for a motor with $N=3$ and $M=1$ but with different
noise levels. The first plot allows one to identify only major
forward and backward transitions at times $t_k$ (marked on the axis)
while the states $[0]_l$ and corresponding substeps are hidden in
the noise. However, the second plot reveals all the transitions and
substeps, so that the $t^0_k$, marking the beginning (or end) of
each full cycle, can be determined. These schematic examples
demonstrate that detectable transitions do not necessarily
correspond to a full biochemical cycle so that a proper statistical
analysis must take account of substeps  hidden in the noise. Notice,
indeed, that the three major transitions identified as forward and
back steps at times $t_3$, $t_4$ and $t_5$ (on the left) are
associated with only a single complete cycle from $t^0_2$ to
$t^0_3$.}
\end{figure*}

To clarify the issues involved, suppose the cycle has four states
($N=4$) and the major transition occurs between states $(2)_l$ and
$(3)_l$ (i.e., $M=2$). Then in stepping time series such as
illustrated in Figs.~1(a) and (b), one can identify all the moments
of time at which the motor leaves state $(2)_l$ and reaches state
$(3)_l$ on moving forwards or when it leaves state $(3)_l$ for state
$(2)_l$ on moving backwards. When successive forward steps are
realized, as illustrated in~Fig.~1(a), one knows that the motor must
pass through the remaining two states, $[0]$ and $(1)$, at some
points between the major transitions [see Fig.~1(a)] because state
$(2)$ cannot otherwise be reached {\it following} state $(3)$ at the
same site $l$. Thus in a sequence of three successive $+$ steps one
can conclude that the middle step is associated with a complete
(forward) cycle. The corresponding observations are equally true for
successive back steps. On the other hand, when the overall stepping
sequence encompasses {\it both} back-steps {\it and} forward-steps,
which is the interesting (and usual) situation [see Fig.~1(b)], it
is {\it impossible}, for example, to be sure that the motor has
completed a full forward cycle when a (detectable) forward step is
followed by a back step; likewise, one cannot tell if a full back
cycle was completed. In such cases, the full-cycle assumption is not
valid.

The full-cycle assumption can be inadequate even when only a run of
forward steps is seen, as in Fig.~1(a), in that back steps may
merely be infrequent. This is somewhat counterintuitive since one
might well argue that each $+$ step does then correspond to a full
cycle. However, if the enzymatic cycle is reversible there is always
the possibility of a completed back step; thus explicit expressions
in terms of the basic rates $\{u_j,\;w_j\}$ will differ when
potentially hidden substeps are allowed for. Never-the-less, as we
demonstrate in Sec.~\ref{sec:level4}, there are various cases in
which the quantitative differences may be small.

To demonstrate further the consequences of different conceivable
interpretations, consider an ($N=3$)-state motor with two possible
substeps. Fig.~1(c) illustrates a stepping series with a relatively
high noise level so that only the major transitions, say
$(1)_l\rightleftharpoons(2)_l$ for $M=1$, at times $t_k$ (with the
corresponding dwell times $\tau_k=t_k-t_{k-1}$) can be measured. On
the other hand, Fig.1(d) shows {\it exactly the same} series of
steps, but with a much lower noise level revealing the previously
obscured small substeps, $[0]_l\rightleftharpoons(1)_l$ and
$(2)_l\rightleftharpoons[0]_{l+1}$. In the latter case, one can
determine the times $t^0_k$ when the motor reaches the bound state
$[0]_l$ for the first time (i.e., when a cycle is completed). And
then one can reliably determine the number $\check{n}_+$ of full
forward, and $\check{n}_-$ of full backward {\it cycles}. In
general, when both forward- and back-steps are present the mean
values of the cycle times $\check{\tau}_i=t^0_i-t^0_{i-1}$ (and so
$\check{\tau}$, $\check{\tau}_+$, and $\check{\tau}_-$) are quite
different from the mean step-to-step dwell times $\tau$, $\tau_+$
and $\tau_-$, that one can obtain from the noisy stepping series in
Fig.1(c). The difference between the splitting probabilities,
$\check{\pi}_{\pm}$ and $\pi_{\pm}$, is even more obvious. For
example in Fig.~1(d) one has {\it only} one back cycle since one
must not consider the major transitions at times $t_3$ and $t_4$ as
indicating full stepping cycles because the motor never actually
reached the next bound state $[0]$: hence from this sequence one
should estimate $\check{\pi}_+\simeq5/6$ and
$\check{\pi}_-\simeq1/6$ and $\check{\pi}_+/\check{\pi}_-\simeq5$.
Conversely in Fig.~1(c) one would count 6 forward and 2 back steps
(or to be more precise, major transitions) leading to the estimates
$\pi_+\simeq3/4$ and $\pi_-\simeq1/4$ so that $\pi_+/\pi_-\simeq3$.

From a mathematical viewpoint, although most of the transitions and
biomechanochemical states remain unseen, there is one bright spot!
Specifically, in light of the basic feature or model assumption
embodied in Eq.~(\ref{BasicModel}), at the instant of time before
the moment, say $t_k$, at which a $+$ step occurs, one can be {\it
sure} the motor was in state ($M$) while in the {\it instant} just
{\it after} $t_k$ the motor is in state ($M+1$); and, likewise, just
{\it before a backward} ($-$) {\it step} the state ($M+1$) is
occupied, while just {\it after} a back-step the state ($M$) is
definitely occupied. Together with the standard Markovian premise of
chemical kinetics, namely, that once in a well defined chemical
state the subsequent departures are independent of the mode of
arrival, this crucial observation enables the systematic calculation
of splitting probabilities and conditional dwell times for general
$N>1$ via the {\it Theory of First Passage Times:} specifically, as
we now explain, we may use the analysis as formulated by van
Kampen~\cite{kampen}.

%***********************************************************************************************************************************
\section{\label{sec:level2}Conditional splitting probabilities and dwell times}

Before undertaking explicit calculations to obtain expressions for
$\pi_+$, $\pi_-$, and $\tau_+$, $\tau_-$, in terms of the $u_i$ and
$w_j$ for general $N$ and $M$, we introduce some further statistical
properties that are straightforward to observe experimentally and
might prove mechanistically informative. At the same time, they
enter naturally in to the first-passage analysis that is presented
in Sec.~\ref{sec:level3}.

In addition to the prior dwell times defined
in~(\ref{PriorDwellTimes}) one may {\it separately} observe {\it
post dwell times} by measuring intervals {\it following after} $+$
or $-$ major steps: we will label the corresponding mean dwell times
$\tau_{+\diamond}$ and $\tau_{-\diamond}$, where the subscript
$\diamond$ is read as `diamond' and denotes, here and below, a $+$
{\it or} a $-$ step. However, such dwell times may be {\it
truncated} by {\it detachments} (or dissociations or disconnections)
in which the motor leaves the track (essentially irreversibly) so
ending a run.  The {\it rates of detachment from states} $(i)$, say
$\delta_i$, can certainly be included in the basic sequential
kinetic model~\cite{fisher99,fisher99a,kolomeisky00}; but in the
{\it first instance} they may be {\it neglected} provided, as we
will suppose, only time intervals between observed $+$ or $-$
mechanical steps are considered. (Their effects, however, would be
significant if dwell times {\it prior to detachments or immediately
following attachments} were considered which might, indeed, prove
informative.)

Neglecting such ``initial'' and ``final'' dwell times (although the
former have been examined by Veigel {\it et al.} for myosin V in
seeking {\it observable} mechanical substeps~\cite{veigel05}) one
may still observe the {\it four distinct conditional mean dwell
times}:
\begin{equation}
\begin{array}{ll}
\!\!\tau_{++}: & \mbox{between two successive forward (+) steps},\\
\!\!\tau_{+-}: & \mbox{between a + step followed by a back-step},\\
\!\!\tau_{-+}: & \mbox{between a back-step followed by a + step},\\
\!\!\tau_{--}: & \mbox{between two successive back-steps},
\end{array}\label{DefPairwiseDwells}
\end{equation}
defined, as in~(\ref{PriorDwellTimes}), in terms of the observed
intervals $\tau^{++}_k=t^{(+)}_k-t^{(+)}_{k-1}$, averaged over
$n_{++}$ pairs of successive $+$ steps, and likewise for $n_{+-}$
pairs of $+$ steps followed by $-$ steps, etc.

Another aspect is to note that for realistic runs of limited length,
deviations of order $1/n$ will arise. Thus, for example, for a run
of length $n=n_++n_-$ {\it starting} with a $+$ step the overall
mean dwell time is given by
\begin{equation}
\tau=[(n_+-1)\tau_++n_-\tau_-]/(n-1),\label{TauCorrection1}
\end{equation}
there being only $(n-1)$ measurable (prior) dwell times
$\tau_l=t_l-t_{l-1}$. Using the definitions (\ref{StepSplProb}) then
yields
\begin{equation}
\tau=\pi_+\tau_++\pi_-\tau_--\frac{\pi_-(\tau_+-\tau_-)}{n-1}.\label{TauCorrection2}
\end{equation}
In fitting asymptotic ($n\gg 1$) expressions to real data from
finite runs such systematic deviations should be recognized. Here,
however, we will neglect such finite-$n$ or end-effects.

To proceed further it is also helpful to introduce the {\it pairwise
step probabilities} $\;\pi_{++}$ and $\pi_{+-}$ defined as the
probability that a $+$ step is followed by a $+$ or, respectively,
by a $-$ step, and likewise, $\pi_{-+}$ and $\pi_{--}$. These then
satisfy
\begin{equation}
\pi_{++}+\pi_{+-}=1\qquad\mbox{and}\qquad\pi_{-+}+\pi_{--}=1.\label{PairwiseSplNorm}
\end{equation}

Again, in a finite run of $n$ steps one can divide the $n-1$
successive pairs into $n_{++}$ of $+$ steps followed by a $+$ step,
and so on, and use $\pi_{++}\approx n_{++}/(n_{++}+n_{+-})$,
$\pi_{+-}\approx n_{+-}/(n_{+-}+n_{++})$, etc. Noting that in a
given run one must have $|n_{+-}-n_{-+}|\leq 1$, and neglecting
finite-$n$ corrections, leads to the valuable relation
\begin{equation}
\pi_{+}\pi_{+-}=\pi_{-}\pi_{-+}.\label{ValuableRelation}
\end{equation}
From this follows the connections
\begin{equation}
\pi_+=1-\pi_-=\pi_{-+}/(\pi_{+-}+\pi_{-+}),\label{StepSplInPairSplPl}
\end{equation}
\begin{equation}
\frac{1}{\pi_+}=1+\frac{\pi_{+-}}{\pi_{-+}}\qquad\mbox{and}\qquad\frac{1}{\pi_-}=1+\frac{\pi_{-+}}{\pi_{+-}}.\label{StepSplInPairSplMi}
\end{equation}
Together with~(\ref{PairwiseSplNorm}) these relations show that the
pair $\pi_{+-}$ and $\pi_{-+}$ or, equivalently, $\pi_{++}$ and
$\pi_{--}$ serve to determine all the back/forward or splitting
probabilities.

It is worthwhile to carry these considerations a stage further by
recognizing the Markovian character of the basic $N$-state
model~(\ref{BasicModel}). Thus, neglecting detachments, the four
division or splitting probabilities $\pi_{++}$, $\pi_{+-}$,
$\pi_{-+}$, and $\pi_{--}$ satisfying~(\ref{PairwiseSplNorm}) can be
regarded as the elements of a $2\!\times\!2$ stepping matrix,
$[\pi_{\alpha\beta}]$, that stochastically determines the
transitions from one (major) step, $+$ or $-$, to the next. By
virtue of the conservation of probability, the largest eigenvalue is
$\;\lambda_0=1;\;$ but the second eigenvalue, which determines the
decay per step of step-step correlations, is just
\begin{eqnarray}
\lambda_1&=&1-\pi_{+-}-\pi_{-+}\nonumber\\
&=&\pi_{++}-\pi_{-+}=\pi_{++}+\pi_{--}-1.\label{Eigenvalue2}
\end{eqnarray}
This vanishes when $\pi_{+-}=\pi_{-+}=\frac{1}{2}$ which corresponds
to $\pi_+=\pi_-$ and hence, to {\it stall conditions} in which the
mean velocity, $V$, vanishes.

Counting arguments similar to those
yielding~(\ref{TauCorrection1})-(\ref{ValuableRelation}) also lead
to relations for the conditional mean dwell times. For completeness
and consistency with later expressions, we utilize the
$+/\!\!-`diamond$' notation introduced before. For the prior dwell
times, we thus find
\begin{eqnarray}
&&\tau_+\equiv\tau_{\diamond+}=\pi_{++}\tau_{++}+\pi_{+-}\tau_{-+},\label{StepDwlInPairDwlPl}\\
&&\tau_-\equiv\tau_{\diamond-}=\pi_{-+}\tau_{+-}+\pi_{--}\tau_{--},\label{StepDwlInPairDwlMi}
\end{eqnarray}
which, in turn, are fully consistent with the relation
(\ref{TotalDwellasSum}) for $\tau$ in terms of $\tau_+$ and $\tau_-$
where we should note
\begin{equation}\label{PiPMDiamond}
    \pi_+\equiv\pi_{\diamond+}\equiv\pi_{+\diamond}\quad\mbox{and}\quad\pi_-\equiv\pi_{\diamond-}\equiv\pi_{-\diamond}.
\end{equation}
Then the {\it post dwell times} likewise satisfy
\begin{eqnarray}
&&\tau_{+\diamond}=\pi_{++}\tau_{++}+\pi_{+-}\tau_{+-},\label{PostDwlInPairDwlPl}\\
&&\tau_{-\diamond}=\pi_{-+}\tau_{-+}+\pi_{--}\tau_{--},\label{PostDwlInPairDwlMi}
\end{eqnarray}
while the overall mean dwell time is given by
\begin{equation}\label{OverallDwlTime}
    \tau\equiv\tau_{\diamond\diamond}=\pi_+\tau_{+\diamond}+\pi_-\tau_{-\diamond}.
\end{equation}

Each of these pairwise fractions and dwell times can be obtained
from the same experimental data (i.e., stepping time series) that
have been used experimentally to obtain the step splitting
probabilities and the prior dwell times in the course of studying
the dynamics of a motor as a function of load and [ATP], etc. But by
observing such further independent statistical parameters one can
test the basic theory more completely and hope to obtain more
reliable and constrained fitting values for the rates determining
the full mechanochemical cycle.

At a more detailed level it is also useful to define
$\;\;n_{i,+}^{\rho\sigma}\;\;$ and $\;\;n_{j,-}^{\rho\sigma}\;\;$
with $\;\;\rho,\sigma=\diamond,+,\;\mbox{or}\;-$, as the mean number
of hidden forward and backward transitions, possibly hidden, from
states $(i)$ and $(j)$, respectively, in the intervals between
(major) steps {\it subject} to the conditions specified by the pair
$(\rho,\sigma)$. If these transitions prove to be detectable, they
can be counted and used in fitting parameters; but if they pertain
to hidden transitions (e.g., the hydrolysis of ATP, etc.), it is of
interest to estimate how often they occur given specific rates. The
appropriate calculations on the basis of the
model~(\ref{BasicModel}) are developed below in
Sec.~\ref{sec:level3e}.

It is appropriate here to mention various hidden-Markov methods,
etc.~\cite{smith01,milescu06,mckinney06,milescu06a}, that have been
derived and employed to locate steps in the presence of noise (and
to fit their amplitudes, or kinetic parameters, etc.). These
approaches require an input stochastic
model~\cite{smith01,milescu06a,foot1}; we believe the present
approach could provide a valuable complement in the extraction of
kinetic parameters from such experimental data since, as we will
see, it reveals the kinds of behavior different models can generate.

%***********************************************************************************************************************************
\section{\label{sec:level3}Explicit calculations}
\subsection{\label{sec:level3a}Formulation and Notation}
The various stepping fractions, dwell times, etc., introduced in
Sec.~\ref{sec:level1} and \ref{sec:level2} can be derived explicitly
in terms of the basic kinetic rates by using van Kampen's analysis
for one-dimensional, nearest-neighbor first-passage
processes~\cite{kampen}. Accordingly, following the basic sequential
model~(\ref{BasicModel}), with the $N$-periodicity
conventions~(\ref{Convention}) for the sequential forward and
backward rates, $u_i$ and $w_j$, we envisage a random walker on a
one-dimensional lattice with sites labeled $i,j=0,\pm1,\pm2,\cdots$,
corresponding, in turn, to the motor states $(i)$, $(j)$, etc.
(again subject to the periodicity convention). If the single major
or observable step per cycle corresponds to the transitions
$(M)\rightleftharpoons(M+1)$ with $M\in[0,N-1]$ we introduce
(following~\cite{kampen}) absorbing boundaries on the left and the
right via
\begin{equation}\label{SpecifyLR}
    L\equiv M\qquad \mbox{and}\qquad R\equiv M+1+N.
\end{equation}

If, for given initial conditions at time $t=0$ (to be selected
below), $q_i(t)$ is the probability that the motor/walker is in
state $(i)$ at time $t$ we may construct the $N\!\times\!N$
transition matrix ${\bf A}=[A_{ij}]$ with elements
\begin{equation}
A_{i,j}=u_j\delta_{i,j+1}+w_j\delta_{i+1,j}-(u_j+w_j)\delta_{i,j},\label{ElemA}
\end{equation}
where $i,j\in[L+1,R-1]=[M+1,M+N]$. Then if
$\textbf{q}^{\textrm{T}}=[q_{M+1},q_{M+2},\cdots,q_{M+N}]$ is the
state vector, the governing rate equations are
\begin{equation}
\frac{d\textbf{q}(t)}{dt}\;=\;\textbf{A}\textbf{q}(t),\label{RateEqnForq}
\end{equation}

This completes the first-passage formulation~\cite{kampen}. Before
proceeding, however, we record some convenient notation for the
various products and sums of the rate constants that enter the
analysis. To that end, our first definition~\cite{foot2}
%\footnote{Note that a
%related but distinct notation was used in Ref.~\cite{kolomeisky00}
%and in A.~B.~Kolomeisky and M.~E.~Fisher, J.~Chem.~Phys.
%\textbf{113}, 10867 (2000): specifically the products
%$\Pi_j^k=\prod_{i=j}^{i=k}(w_i/u_i)\equiv\Gamma_{j-1,k-j+1}$ were
%defined along with
%$\Pi_1^N\equiv\Pi_{0}^{N-1}=1/\Gamma=\Gamma_{l,N}\equiv\Gamma_N$,
%where $\Gamma_{l,m}$ is defined in~(\ref{DefGamma}).}
is of the $(m\!\geqslant\!1)$-term product
\begin{equation}
\Gamma_{l,m}=\prod_{j=1}^{m}\frac{w_{l+j}}{u_{l+j}},\label{DefGamma}
\end{equation}
which, by periodicity, is invariant under $l\Rightarrow l\pm N$.
Likewise, the $N$-term product $\Gamma_{l,N}$ is independent of $l$
yielding, specifically,
\begin{equation}
\Gamma_N=\Gamma_{l,N}=\prod_{j=0}^{N-1}\frac{w_j}{u_j},\label{GammaN}
\end{equation}
\cite{foot2}. Then for all $l=0,\pm1,\pm2,\ldots$ a central role
will be played by the $(n-1)$-term sum
\begin{equation}
\Delta_{l,n}=\sum_{m=1}^{n-1}\Gamma_{l,m}\qquad(n\geqslant1),\label{DefDelta}
\end{equation}
where, for the empty sum, we set $\Delta_{l,1}\equiv0$. Indeed,
these sums appear in previous
analyses~\cite{fisher99,fisher99a,kolomeisky00,foot2} via
``renormalized'' inverse forward rates (or transition times)
\begin{equation}
r_l=u_l^{-1}(1+\Delta_{l,N}).\label{TransitionTimes}
\end{equation}
Specifically, these enter into the
expression~\cite{fisher99,fisher99a,kolomeisky00}
\begin{equation}
V/d=(1-\Gamma_N)\left/\mbox{$\sum_{l=0}^{N-1}r_l$}\right.\label{Velocity}
\end{equation}
for the {\it mean velocity} $V$, which we recall here for
convenience of reference. (Note that $d$ is the mean spacing of
sites along the track.) One sees directly from this that {\it stall
conditions}, i.e., $V=0$, are determined by $\Gamma_N({u_i,w_j})=1$.
The situation near stall will be a major focus for our discussions
in Sec.~\ref{sec:level4}.

The analysis of van Kampen~\cite{kampen} may now be put to work.
Readers uninterested in the details may skip to the next section or
peruse the main results, namely,
(\ref{SplProbPlPl})-(\ref{StepSplProbMi}) for $\pi_{++}$, etc.,
(\ref{DwelPlPl})-(\ref{DwelMiMi}) for $\tau_{++}$, etc., and
(\ref{DwelPl})-(\ref{DwelTot}) for $\tau_{+}$, $\tau_{-}$ and
$\tau$.

%---------------------------------------------------------------------------------

\subsection{\label{sec:level3b}Pairwise step splitting probabilities}
To proceed, let $\pi_k^L$ be the total probability that a motor
starting at $t=0$ in state $(k)$ with $L<k<R$, so that
$q_i(0)=\delta_{ik}$, eventually reaches the left absorbing state
$(L)$ for the first time, i.e., without having been absorbed at
sites $L$ {\it or} $R$. A moment's thought confirms
\begin{equation}
    \pi_k^L=w_{L+1}\int_0^{\infty}q_{L+1}(t)dt;\label{SplGenB}
\end{equation}
likewise, for reaching the boundary state $(R)$ for the first time,
one has
\begin{equation}
    \pi_k^R=u_{R-1}\int_0^{\infty}q_{R-1}(t)dt.\label{SplGenF}
\end{equation}
Then we may appeal to Ref.~\cite{kampen}, Ch.XII, Eq.(2.8) which,
using the notation (\ref{DefGamma}), states
\begin{equation}
\pi_k^L=1-\pi_k^R=\frac{\sum_{m=k-L}^{R-L-1}\Gamma_{L,m}}{1+\sum_{m=1}^{R-L-1}\Gamma_{L,m}}.\label{vanKampen28}
\end{equation}

For $L$ and $R$ we have (\ref{SpecifyLR}). If the motor starts just
after a (major) forward or $+$ step it is in the initial state
$(k)=(M+1)$; then $\pi_k^L$ and $\pi_k^R$ correspond, respectively,
to $\pi_{+-}$ and $\pi_{++}$. Conversely, just after a (major) back
or $-$ step the motor is in state $(M)$ which, by periodicity is
equivalent to starting $k=M+N$; then we may identify $\pi_k^L$ and
$\pi_k^R$ with $\pi_{--}$ and $\pi_{-+}$, respectively. On using the
notation (\ref{GammaN}) and (\ref{DefDelta}) we thus obtain
\begin{eqnarray}
&&\pi_{++}(M,N)=1-\pi_{+-}=\frac{1}{1+\Gamma_N+\Delta_{M,N}},\label{SplProbPlPl}\\
&&\pi_{-+}(M,N)=1-\pi_{--}=\frac{1+\Delta_{M,N}}{1+\Gamma_N+\Delta_{M,N}}.\label{SplProbMiPl}
\end{eqnarray}

It is interesting to note the cross-relation
\begin{equation}\label{CrossRelation}
    \pi_{--}=\Gamma_N\pi_{++}.
\end{equation}
At stall this implies $\pi_{++}=\pi_{--}$ and $\pi_{+-}=\pi_{-+}$
which reflects the expected $+\!/\!-$ symmetry. The results
(\ref{SplProbPlPl}) and (\ref{SplProbMiPl}) also yield, via the
identities (\ref{StepSplInPairSplPl}) and
(\ref{StepSplInPairSplMi}), the explicit step fraction expressions
\begin{eqnarray}
&&\pi_{+}=1-\pi_{-}=\frac{1+\Delta_{M,N}}{1+\Gamma_N+2\,\Delta_{M,N}},\label{StepSplProbPl}\\
&&\pi_{-}=\frac{\Gamma_N+\Delta_{M,N}}{1+\Gamma_N+2\,\Delta_{M,N}}.\label{StepSplProbMi}
\end{eqnarray}

%---------------------------------------------------------------------------------

\subsection{\label{sec:level3c}Pairwise and prior dwell times}

Now, following van Kampen~\cite{kampen}, Ch.XII, Eqs.(1.7)-(1.8),
the {\it conditional mean first-passage times} for arriving either
at the left or right absorbing boundaries starting from a state
point $k\in[L+1,R-1]=[M+1,M+N]$ as before, are
\begin{equation}
\tau_k^L=\frac{\int_0^{\infty}tq_{L+1}(t)dt}{\int_0^{\infty}q_{L+1}(t)dt},
\qquad
\tau_k^R=\frac{\int_0^{\infty}tq_{R-1}(t)dt}{\int_0^{\infty}q_{R-1}(t)dt},\label{DefDwelGen}
\end{equation}
where the $q_i(t)$ are the solutions of (\ref{RateEqnForq}) subject,
for our purposes, to the two, alternative initial conditions,
\begin{equation}\label{InitialConditions}
    q_i^+(0)=\delta_{i,M+1}\quad\mbox{and}\quad
    q_i^-(0)=\delta_{i,M+N},
\end{equation}
as discussed in deriving (\ref{SplProbPlPl}) and
(\ref{SplProbMiPl}).

To obtain the pairwise conditional mean dwell times we proceed in
two steps. First, we integrate the kinetic equations
(\ref{RateEqnForq}) over all times recognizing that the $q_j(t)$
approach zero exponentially fast for all $j\in[M+1,M+N]$, since the
walker must eventually be absorbed at either $L$ or $R$. This yields
\begin{equation}
-\textbf{q}^{\pm}(0)=\textbf{A}\textbf{T}^{\pm},\label{RateEqnForQ}
\end{equation}
where the superscripts identify the alternative initial conditions
(\ref{InitialConditions}). The elements of the vector
$\textbf{T}^{\pm}$ have the dimensions of time and are given simply
by
\begin{equation}
T_{k}^{\pm}=\int_0^{\infty}q^{\pm}_k(t)dt.\label{DefQ}
\end{equation}
By the definitions (\ref{SplGenB}) and (\ref{SplGenF}) with
(\ref{SpecifyLR}), we also have the relations
\begin{eqnarray}
T_{M+1}^{+}=\pi_{+-}/w_{M+1},&\;& T_{M+1}^{-}=\pi_{--}/w_{M+1},\label{TatBoundL}\\
T_{M+N}^{+}=\pi_{++}/u_{M+N},&\;&
T_{M+N}^{-}=\pi_{-+}/u_{M+N}.\label{TatBoundR}
\end{eqnarray}

Since $\textbf{A}$ is a tridiagonal matrix the equations
(\ref{RateEqnForQ}) can be inverted recursively to obtain
\begin{eqnarray}
T_j^+&=&\pi_{++}\frac{\Gamma_N+\Delta_{M,N}-\Delta_{M,j-M}}{u_j\Gamma_{M,j-M}},\label{SolutionQPl}\\
T_j^-&=&\pi_{--}\frac{1+\Delta_{M,j-M}}{u_j\Gamma_{M,j-M}}.\label{SolutionQMi}
\end{eqnarray}
One general approach to this inversion can be found in
Ref.~\cite{kampen}, Ch.XII, Sec.2. On recalling
$\Delta_{l,1}\equiv0$, one may check that these solutions verify the
relations (\ref{TatBoundL}) and (\ref{TatBoundR}).

The next step is to multiply (\ref{RateEqnForq}) by $t$ and again
integrate over all time which yields
\begin{equation}
-\textbf{T}^{\pm}=\textbf{A}\textbf{S}^{\pm},\label{RateEqnForT}
\end{equation}
where, with the same superscript conventions, etc., we have
\begin{equation}
S_{k}^{\pm}=\int_0^{\infty}tq^{\pm}_k(t)dt.\label{DefT}
\end{equation}
From  (\ref{DefDwelGen}) and following the arguments above
(\ref{SplProbPlPl}) and (\ref{SplProbMiPl}), we can now make the
identifications
\begin{eqnarray}
  \tau_k^{L+} &=& \tau_{+-}=S_{M+1}^{+}/T_{M+1}^{+},\label{TAUboundL} \\
  \tau_k^{R-} &=& \tau_{-+}=S_{M+N}^{-}/T_{M+N}^{-},\label{TAUboundR}
\end{eqnarray}
and similarly for $\tau_{--}$ and $\tau_{++}$. Inverting
(\ref{RateEqnForQ}) finally leads to the basic pairwise dwell time
expressions
\begin{eqnarray}
\tau_{++}&=&\sum_{j=M+1}^{M+N}T_j^+(1+\Delta_{M,j-M}),\label{DwelPlPl}\\
\tau_{-+}&=&\sum_{j=M+1}^{M+N}T_j^-\frac{1+\Delta_{M,j-M}}{1+\Delta_{M,N}},\label{DwelMiPl}\\
\tau_{+-}&=&\sum_{j=M+1}^{M+N}T_j^+\frac{\Gamma_N+\Delta_{M,N}-\Delta_{M,j-M}}{\Gamma_N+\Delta_{M,N}},\label{DwelPlMi}\\
\tau_{--}&=&\sum_{j=M+1}^{M+N}T_j^-\frac{\Gamma_N+\Delta_{M,N}-\Delta_{M,j-M}}{\Gamma_N}.\label{DwelMiMi}
\end{eqnarray}

By using (\ref{SplProbPlPl}) and (\ref{SplProbMiPl}) for $\pi_{++}$
and $\pi_{--}$ in (\ref{SolutionQPl}) and (\ref{SolutionQMi}) we may
establish the unanticipated, general equality
\begin{equation}\label{TauIdentity}
    \tau_{++}=\tau_{--},\qquad\mbox{all}\;M,\;N.
\end{equation}

En route to the prior dwell times $\tau_+$ and $\tau_-$ it is
convenient to introduce
\begin{equation}
T_j=\pi_+T_j^++\pi_-T_j^-=\frac{(1+\Delta_{j,N})/u_j}{1+\Gamma_N+2\;\Delta_{M,N}},\label{ReDefQ}
\end{equation}
where we have used (\ref{StepSplProbPl}) and (\ref{StepSplProbMi})
and may recall that the numerator is the inverse rate defined in
(\ref{TransitionTimes}) and used in the past. Note also that by
virtue of the periodicity we have $T_{j\pm N}\equiv T_j$ and,
likewise, for the $T_j^{\pm}$. Then, by utilizing
(\ref{StepDwlInPairDwlPl}), (\ref{StepDwlInPairDwlMi}),
(\ref{SplProbPlPl}), and (\ref{SplProbMiPl}) we obtain the prior
dwell times in the form
\begin{align}
  &\tau_{+}\equiv\tau_{\diamond+}=\frac{1}{\pi_+}\sum_{j=M+1}^{M+N}T_j\frac{1+\Delta_{M,j-M}}{1+\Gamma_N+\Delta_{M,N}},\label{DwelPl}\\
  &\tau_{-}\equiv\tau_{\diamond-}=\frac{1}{\pi_-}\sum_{j=M+1}^{M+N}T_j\frac{\Gamma_N+\Delta_{M,N}-\Delta_{M,j-M}}{1+\Gamma_N+\Delta_{M,N}}.\label{DwelMi}
\end{align}
Finally, with the aid of (\ref{TotalDwellasSum}), the overall mean
dwell time is simply
\begin{equation}
    \tau=\sum_{j=M+1}^{M+N}T_j\equiv\sum_{i=0}^{N-1}T_i.\label{DwelTot}
\end{equation}

Although, we have introduced the various step fractions via the
pairwise fractions $\pi_{++}$, etc., this was not a necessary move
from the mathematical point of view. Indeed, one can the
results~(\ref{StepSplProbPl}) and (\ref{StepSplProbMi}) for $\pi_+$
and $\pi_-$, and the present expressions for $\tau_+$, $\tau_-$ and
$\tau$, directly by solving the basic rate
equations~(\ref{RateEqnForq}) with the initial conditions
\begin{equation}
q_k(0)=\pi_+\delta_{k,M+1}+(1-\pi_+)\delta_{k,M+N},
\end{equation}
together with the relations~(\ref{RateEqnForQ}) and
(\ref{RateEqnForT}) and appropriate boundary conditions. By this
route one need not mention the pairwise splitting probabilities or
pairwise dwell times. Nevertheless, the pairwise stepping fractions
and dwell times can be useful in data analysis and to test theory,
since they represent additional force and [ATP] dependent parameters
that can be measured without significant extra experimental effort.

%---------------------------------------------------------------------------------

\subsection{\label{sec:level3d}Individual and post dwell times}

Now the form~(\ref{DwelTot}) for $\tau$ strongly suggests that
$T_j\equiv T_j^{\diamond\diamond}$ is actually the overall {\it
individual mean dwell time} spent in state $(j)$ irrespective, as
indicated by the use of the $+\!/\!-$ or diamond notation [see
(\ref{StepDwlInPairDwlPl})-(\ref{OverallDwlTime})] of the stepping
sequence. This conclusion is, indeed, justified since it follows
from~(\ref{DefQ}) that we may identify $T_j^+=T_j^{+\diamond}$ and
$T_j^-=T_j^{-\diamond}$ as {\it individual post} $+$ and $-$ step
{\it mean dwell times} in state $(j)$, respectively. Consequently,
the mean overall {\it post dwell times}, are given by
\begin{equation}
\tau_{+\diamond}=\sum_{i=0}^{N-1}T_i^{+\diamond}\quad\mbox{and}\quad\tau_{-\diamond}=\sum_{i=0}^{N-1}T_i^{-\diamond}.\label{PostDwelAsSumT}
\end{equation}
From these results one may verify that~(\ref{OverallDwlTime}) is
satisfied.

Likewise, we anticipate relations like
\begin{equation}
\tau_{++}=\sum_{i=0}^{N-1}T_i^{++},\label{TauPlPlAsSum}
\end{equation}
 etc., and, hence, from
(\ref{DwelPlPl}) and (\ref{DwelMiMi}) we surmise that the {\it
conditional individual state dwell times} are
\begin{align}
 T_j^{++}&=T_j^{--} \nonumber\\
 &=\frac{(1+\Delta_{M,j-M})(\Gamma_N+\Delta_{M,N}-\Delta_{M,j-M})}{u_j\;\Gamma_{M,j-M}\;(1+\Gamma_N+\Delta_{M,N})},\label{DefTPlPl}
\end{align}
while (\ref{DwelPlMi}) and (\ref{DwelMiPl}) yield
\begin{align}
   &\!\!T_j^{-+}=\frac{\Gamma_N(1+\Delta_{M,j-M})^2}{u_j\Gamma_{M,j-M}(1+\Delta_{M,N})(1+\Gamma_N+\Delta_{M,N})},\label{DefTMiPl}\\
   &\!\!T_j^{+-}=\frac{(\Gamma_N+\Delta_{M,N}-\Delta_{M,j-M})^2}{u_j\Gamma_{M,j-M}(\Gamma_N+\Delta_{M,N})(1+\Gamma_N+\Delta_{M,N})}.\label{DefTPlMi}
\end{align}
In terms of these we can define the {\it prior individual} (or {\it
partial}) {\it dwell times} via
\begin{eqnarray}
   T_j^{\diamond+}&=&\pi_{++}T_j^{++}+\pi_{+-}T_j^{-+},\nonumber\\
   &=&\frac{r_j(1+\Delta_{M,j-M})}{(1+\Delta_{M,N})(1+\Gamma_N+\Delta_{M,N})},\label{DefPriorTPl}
\end{eqnarray}
\begin{eqnarray}
   T_j^{\diamond-}&=&\pi_{-+}T_j^{+-}+\pi_{--}T_j^{--},\nonumber\\
   &=&\frac{r_j(\Gamma_N+\Delta_{M,N}-\Delta_{M,j-M})}{(\Gamma_N+\Delta_{M,N})(1+\Gamma_N+\Delta_{M,N})},\label{DefPriorTMi}
\end{eqnarray}
where the $r_j$ are defined in~(\ref{TransitionTimes}). Hence one
can check the expressions for the mean overall {\it prior} dwell
times, $\tau_{\diamond+}$ and $\tau_{\diamond-}$, given
in~(\ref{DwelPl}) and~(\ref{DwelMi}).

Evidently, the analysis presented does not fully justify the
inferences regarding (\ref{DefTPlPl})-(\ref{DefPriorTMi}). However,
these expressions have been checked by direct computation for $N=2$
(as recorded in Appendix A) and the various cross-checks for general
$M$ and $N$ also serve as validation. However, a complete
justification requires a more elaborate calculation that we hope to
present in the future. By the same route one can derive the mean
conditional counts, $n_{j,+}^{\rho\sigma}$ and
$n_{j,-}^{\rho\sigma}$ [see after
Eqs.~(\ref{PostDwlInPairDwlPl})-(\ref{OverallDwlTime})] of the {\it
hidden substeps} as we now proceed to demonstrate. Corresponding
results for $N=2$ are also presented in Appendix A.

%---------------------------------------------------------------------------------

\subsection{\label{sec:level3e}Counting the hidden substeps}

As touched on briefly in the penultimate paragraph of
Sec.~\ref{sec:level2}, it is surely of interest in light of our
basic premise to estimate for a particular model how many hidden
substeps actually arise on average in the typical intervals
$\tau\equiv\tau_{\diamond\diamond}$, $\tau_{+\diamond}$,
$\tau_{++}$, etc., between specified successive observable steps,
i.e., major transitions between states $(M)$ and $(M+1)$. Granted
the results obtained in the previous subsection for the
$T_j^{\rho\sigma}$, where henceforth, $(\rho,\sigma)$ runs through
the nine combinations
\begin{equation}\label{9combinations}
    \{\rho,\sigma\}=\{\diamond\diamond;+\diamond,-\diamond,\diamond+,\diamond-;++,+-,-+,--\},
\end{equation}
this is a reasonably straightforward exercise.

Notice, first, that the mean dwell time in a given state $(j)$, say
$\bar{\tau}_j$, is, by virtue of the Markovian character of the
relevant biochemical reactions {\it independent} of whether the
state was reached from state $(j-1)$ or $(j+1)$ {\it and} of whether
the motor departs to states $(j+1)$ or $(j-1)$: formally, we may
write
\begin{eqnarray}
    \bar{\tau}_j=(u_j+w_j)^{-1}&=&\tau_j^{(-)\rightarrow(+)}=\tau_j^{(-)\rightarrow(-)}\nonumber\\
                               &=&\tau_j^{(+)\rightarrow(+)}=\tau_j^{(+)\rightarrow(-)},\label{tauBAR}
\end{eqnarray}
where the superscripts have an obvious interpretation.

\begin{figure*}
\includegraphics[angle=0, scale=0.70]{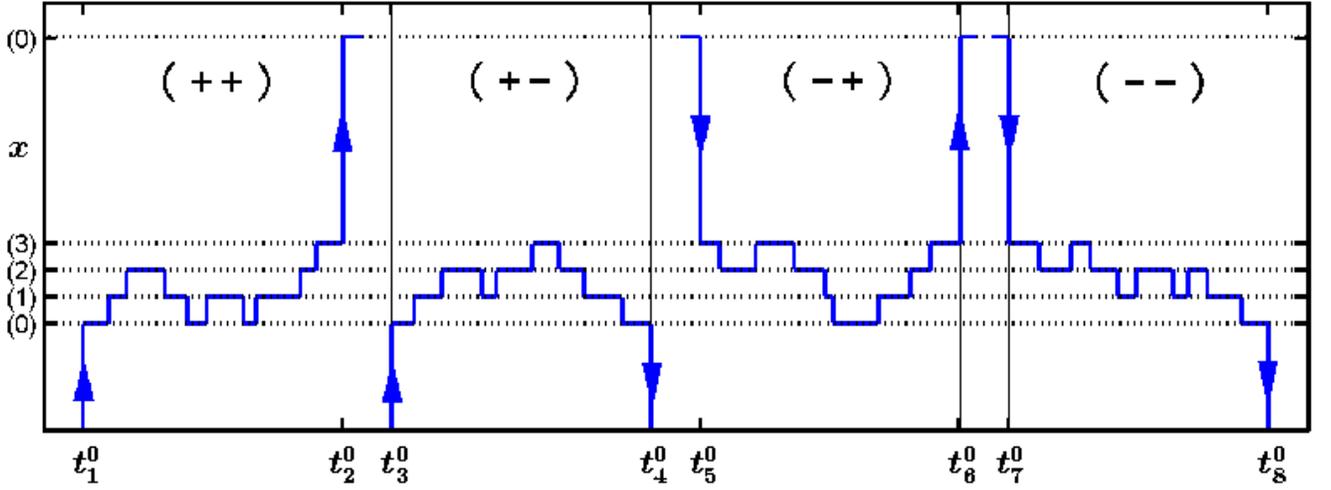}
\caption{\label{fig2:wide}Schematic plots illustrating substeps and
their counting rules (see text) in the $N=4$ case with $M=3$ for
$(++)$, $(+-)$, $(-+)$, and $(--)$ stepping intervals. Thus, for
example, from the first plot one confirms
$n^{++}_{0,+}=n^{++}_{1,-}+1$ and, correspondingly, that the motor
visited the state $[0]$ three times in this realization; and,
likewise, $n^{++}_{1,+}+n^{++}_{1,-}=n^{++}_{0,+}+n^{++}_{2,-}$ is,
here, equal to $4$ so that, correspondingly, the motor spent time in
the substate $(1)$ on four occasions before undergoing the forward
major transition $(3)\!\rightarrow\!(4\equiv0)$. Similarly, the
other plots serve to establish the frequency
relations~(\ref{DefFrequencyJ1})-(\ref{FrequencyRMiMi}). }
\end{figure*}

Then, if $n_{j\rightarrow j+1}$ and $n_{j\rightarrow j-1}$ are the
number of transitions, i.e., substeps, forwards or backwards,
respectively, from state $(j)$ in an interval between observable
steps we desire the conditional mean values
\begin{equation}\label{DefSubnumber}
    n_{j,+}^{\rho\sigma}=\left\langle n_{j\rightarrow j+1}\right\rangle_{\rho\sigma}\quad\mbox{and}\quad n_{j,-}^{\rho\sigma}=\left\langle n_{j\rightarrow j-1}\right\rangle_{\rho\sigma}.
\end{equation}
In terms of these we will have for the overall mean numbers of $+$
or $-$ hidden transitions per step-to-step interval
\begin{eqnarray}
    \bar{n}_{+}^{\rho\sigma}&=&\frac{1}{N-1}\sum_{j=M+1}^{M+N-1}n_{j,+}^{\rho\sigma},\label{TotSubnumberPl}\\
    \bar{n}_{-}^{\rho\sigma}&=&\frac{1}{N-1}\sum_{j=M+2}^{M+N}n_{j,-}^{\rho\sigma},\label{TotSubnumberMi}
\end{eqnarray}
for all nine pairs $(\rho,\sigma)$. As a moments thought reveals,
the limits on the summations here must be carefully set: thus, after
a forward step to state $(M+1)$ any putative forward hidden substep
transformation from state $(M+N)$ would represent a full {\it
observable} $+$ step (i.e.,  a major transition) and so is to be
excluded from the sum of {\it substeps}; equally, any back
transition from state $(M+1)$ would represent an observable
(back)step whereas back {\it sub}steps  from state $(M+N)$, prior to
the final forward step, {\it are} to be counted: See, for example,
Fig.~\ref{fig2:wide} which can be regarded as a noise-free version
of Fig.~\ref{fig:wide}(d) [but for an $(N=4)$-state model].

Now consider $T_j^{\rho\sigma}$, the mean time spent during a
$(\rho,\sigma)$ step interval in a state $(j)$ that is {\it neither}
an initial {\it nor} a final state of a major transition, so that
$j\neq M,\;M+1$. On average this state will be visited on
$T_j^{\rho\sigma}\left/\bar{\tau}_j\right.=(u_j+w_j)T_j^{\rho\sigma}$
separate occasions during the interval; thus it entails
$n_{j,+}^{\rho\sigma}$ forward substep {\it departures} and
$n_{j,-}^{\rho\sigma}$ back substep departures. Equally, it entails,
on average, $n_{j-1,+}^{\rho\sigma}$ forward substep {\it arrivals}
and $n_{j+1,-}^{\rho\sigma}$ back substep arrivals. As a result we
have the {\it frequency relations}
\begin{equation}\label{DefFrequencyJ1}
    \omega_j^{\rho\sigma}\equiv(u_j+w_j)T_j^{\rho\sigma}=n_{j,+}^{\rho\sigma}+n_{j,-}^{\rho\sigma},
\end{equation}
for $j\in[M+2,M+N-1]$ and, similarly,
\begin{equation}\label{DefFrequencyJ2}
    \omega_j^{\rho\sigma}=n_{j-1,+}^{\rho\sigma}+n_{j+1,-}^{\rho\sigma},
\end{equation}
where the units of $\omega_j$ may be regarded as substeps per
interval.

For the reasons explained after~(\ref{TotSubnumberMi}) and
illustrated in Fig.~\ref{fig2:wide} the `boundary states' $j=M+1$
and $j=M+N$, require special consideration. As noted, one cannot
have {\it sub}steps that go backwards from state $(M+1)$ and, in the
case of a prior forward step, the certain incoming arrival must be
included in the individual mean dwell time. Thus we are led to the
four boundary frequency relations
\begin{eqnarray}
  \omega_{M+1}^{++} &=& n_{M+1,+}^{++} = n_{M+2,-}^{++}+1,\label{FrequencyLPlPl}\\
  \omega_{M+1}^{+-} &=& n_{M+1,+}^{+-}+1 = n_{M+2,-}^{+-}+1,\label{FrequencyLPlMi}\\
  \omega_{M+1}^{-+} &=& n_{M+1,+}^{-+} = n_{M+2,-}^{+-},\label{FrequencyLMiPl}\\
  \omega_{M+1}^{--} &=& n_{M+1,+}^{--}+1 = n_{M+2,-}^{--}.\label{FrequencyLMiMi}
\end{eqnarray}
Complementary arguments apply for the opposite boundary state
$(j)\equiv(M+N)$, yielding
\begin{eqnarray}
  \omega_{M+N}^{++} &=& n_{M+N,-}^{++}+1 = n_{M+N-1,+}^{++},\label{FrequencyRPlPl}\\
  \omega_{M+N}^{+-} &=& n_{M+N,-}^{+-} = n_{M+N-1,+}^{+-},\label{FrequencyRPlMi}\\
  \omega_{M+N}^{-+} &=& n_{M+N,-}^{-+}+1 = n_{M+N-1,+}^{-+}+1,\label{FrequencyRMiPl}\\
  \omega_{M+N}^{--} &=& n_{M+N,-}^{--} = n_{M+N-1,+}^{--}+1.\label{FrequencyRMiMi}
\end{eqnarray}

The frequency relations~(\ref{DefFrequencyJ1})
and~(\ref{DefFrequencyJ2}) with the boundary
relations~(\ref{FrequencyLPlPl})-(\ref{FrequencyRMiMi}) constitute,
together with the definition
$\omega_j^{\rho\sigma}=(u_j+w_j)T_j^{\rho\sigma}$
in~(\ref{DefFrequencyJ1}), a complete set from which we may derive
explicit general expressions for all the $n_{j,+}^{\rho\sigma}$ and
$n_{j,-}^{\rho\sigma}$. To proceed, we note first that the purely
counting arguments involving $\pi_+$, $\pi_-$, and $\pi_{++}$,
$\pi_{--}$, etc., that led to the `reduced'
relations~(\ref{StepDwlInPairDwlPl},\ref{StepDwlInPairDwlMi})
and~(\ref{PostDwlInPairDwlPl},\ref{PostDwlInPairDwlMi}), apply
equally to $\omega_j^{++}$, $\omega_j^{+-}$, etc. Accordingly, we
obtain the {\it post} boundary frequency relations
\begin{eqnarray}
  \omega_{M+1}^{+\diamond} &=& n_{M+2,-}^{+\diamond}+1 = n_{M+1,+}^{+\diamond}+\pi_{+-},\label{FrequencyLPlD}\\
  \omega_{M+N}^{+\diamond} &=& n_{M+N-1,+}^{+\diamond} = n_{M+N,-}^{+\diamond}+\pi_{++},\label{FrequencyRPlD}\\
  \omega_{M+1}^{-\diamond} &=& n_{M+2,-}^{-\diamond} = n_{M+1,+}^{-\diamond}+\pi_{--},\label{FrequencyLMiD}\\
  \omega_{M+N}^{-\diamond} &=& n_{M+N-1,+}^{-\diamond}+1 = n_{M+N,-}^{-\diamond}+\pi_{-+},\label{FrequencyRMiD}
\end{eqnarray}
while the {\it prior} relations are
\begin{eqnarray}
  \omega_{M+1}^{\diamond+} &=& n_{M+1,+}^{\diamond+} = n_{M+2,-}^{\diamond+}+\pi_{++},\label{FrequencyLDPl}\\
  \omega_{M+N}^{\diamond+} &=& n_{M+N,-}^{\diamond+}+1 = n_{M+N-1,+}^{\diamond+}+\pi_{+-},\label{FrequencyRDPl}\\
  \omega_{M+1}^{\diamond-} &=& n_{M+1,+}^{\diamond-}+1 = n_{M+2,-}^{\diamond-}+\pi_{-+},\label{FrequencyLDMi}\\
  \omega_{M+N}^{\diamond-} &=& n_{M+N,-}^{\diamond-} = n_{M+N-1,+}^{\diamond-}+\pi_{--},\label{FrequencyRDMi}
\end{eqnarray}
Finally, in analogy to~(\ref{OverallDwlTime}), we obtain
\begin{eqnarray}
  \omega_{M+1}^{\diamond\diamond} &=& n_{M+2,-}^{\diamond\diamond}+\pi_+ = n_{M+1,+}^{\diamond\diamond}+\pi_-,\label{FrequencyLDD}\\
  \omega_{M+N}^{\diamond\diamond} &=& n_{M+N-1,+}^{\diamond\diamond}+\pi_- = n_{M+N,-}^{\diamond\diamond}+\pi_+.\label{FrequencyRDD}
\end{eqnarray}

Now~(\ref{FrequencyLDD}) clearly implies the result
\begin{equation}\label{nM1DD}
    n_{M+1,+}^{\diamond\diamond}=\omega_{M+1}^{\diamond\diamond}-\pi_-=(u_j+w_j)T_{M+1}^{\diamond\diamond}-\pi_-,
\end{equation}
where $\pi_-$ and $T_{M+1}^{\diamond\diamond}\equiv
T_{M+1}^{\diamond\diamond}\!\left(\{u_j,w_j\};M,N\right)$ are given
explicitly in~(\ref{StepSplProbMi}) and~(\ref{ReDefQ}). A similar
result follows from~(\ref{FrequencyLDD}) for
$n_{M+2,-}^{\diamond\diamond}$. But by appealing
to~(\ref{DefFrequencyJ2}) for $j=M+2$ we also obtain
\begin{equation}\label{nM3DD}
    n_{M+3,-}^{\diamond\diamond}=\omega_{M+2}^{\diamond\diamond}-n_{M+1,+}^{\diamond\diamond},
\end{equation}
which, with the aid of~(\ref{nM1DD}), yields an explicit expression
in terms of $T_{M+1}$ and $T_{M+2}$. Substituting $j=M+3$
in~(\ref{DefFrequencyJ1}) then leads to an expression for
$n_{M+3,+}^{\diamond\diamond}$. By proceeding recursively in this
fashion we eventually obtain, for $k=1,2,\cdots$, the general
expressions
%\begin{eqnarray}
\begin{align}
  &n_{M+2k,-}^{\diamond\diamond}=\sum_{j=M+1}^{M+2k-1}(-)^{j-M-1}(u_j+w_j)T_j^{\diamond\diamond}-\pi_+,\label{nDDeven}\\
  &n_{M+2k+1,-}^{\diamond\diamond}=\sum_{j=M+1}^{M+2k}(-)^{j-M}(u_j+w_j)T_j^{\diamond\diamond}+\pi_-,\label{nDDodd}
\end{align}
%\end{eqnarray}
for the average number of hidden {\it back-substeps} from the $N-2$
hidden intermediate states $(M+2)$ to $(M+N-1)$ and from the
pre-step state $(M+N)\equiv(M)$. We may also note the sum rule
\begin{equation}\label{SumRule}
    \sum_{l=0}^{N-1}(-)^l(u_l+w_l)T^{\diamond\diamond}_l=\left\{\begin{array}{cccc}
        0 & \mbox{for} & N & \mbox{even},\\
        1 & \mbox{for} & N & \mbox{odd},
        \end{array}\right.
\end{equation}
and the special relations $\pi_+=u_{M+N}T_{M+N}^{\diamond\diamond}$
and $\pi_-=w_{M+1}T_{M+1}^{\diamond\diamond}$, so that
$n_{M+1,+}^{\diamond\diamond}=u_{M+1}T_{M+1}^{\diamond\diamond}$ and
$n_{M+N,-}^{\diamond\diamond}=w_{M+N}T_{M+N}^{\diamond\diamond}$.

More generally~(\ref{nDDeven}) and~(\ref{nDDodd}) can be extended to
arbitrary $(\rho,\sigma)$ if, following the
sequence~(\ref{9combinations}), $\pi_+$ and $\pi_-$ are replaced by
\begin{eqnarray}
  &&\tilde{\pi}_+=\{\pi_+;\;1,0,\pi_{++},\pi_{-+};\;1,1,0,0\},\label{PiSetPl}\\
  &&\tilde{\pi}_-=\{\pi_-;\;\pi_{+-},\pi_{--},0,1;\;0,1,0,1\},\label{PiSetMi}
\end{eqnarray}
respectively. Finally, one finds that the mean number of {\it
forward substeps} from the $N-2$ hidden intermediate states and the
post-step state $(M+1)$ in specified intervals can be written
\begin{align}
  &\!n_{M+2k,+}^{\rho\sigma}=\!\!\!\sum_{j=M+1}^{M+2k}\!\!\!(-)^{j-M}(u_j+w_j)T_j^{\rho\sigma}+\tilde{\pi}_+,\label{nRSeven}\\
  &\!n_{M+2k-1,+}^{\rho\sigma}=\!\!\!\sum_{j=M+1}^{M+2k-1}\!\!\!(-)^{j-M-1}(u_j+w_j)T_j^{\rho\sigma}-\tilde{\pi}_-,\label{nRSodd}
\end{align}
for $k=1,2,\cdots$. Some examples of these various expressions for
small $N$ are listed in Appendix~A.

%***********************************************************************************************************************************
\section{\label{sec:level4}Discussion and illustrations}
First, it is appropriate to look more closely at the difference
between our present results
[see~(\ref{StepSplProbPl}),(\ref{StepSplProbMi}), and
(\ref{DwelPl})-(\ref{DwelTot})] and those of the previously
published analysis~\cite{kolomeisky03,kolomeisky05}. As in
Sec.~\ref{sec:level1} [after~(\ref{MajorMechStep})] we use a
ha$\check{\textrm{c}}$ek to distinguish the results that presuppose
the completion of a full enzymatic cycle between all pairs of
successive major, i.e., observable steps. In the present notation
one then has~\cite{kolomeisky05}
\begin{equation}
   \check{\pi}_+=\frac{1}{1+\Gamma_N},\qquad\check{\pi}_-=\frac{\Gamma_N}{1+\Gamma_N},\label{SplOld}
\end{equation}
for the forward/back fractions while the mean, prior, and post dwell
times are all given by
\begin{equation}
   \check{\tau}=\check{\tau}_+=\check{\tau}_-=\sum_{n=0}^{N-1}r_n\left/(1+\Gamma_N)\right.,\label{DwelOld}
\end{equation}
where $r_n$ is defined in~(\ref{TransitionTimes}). On the other
hand, when allowing for hidden substeps, these three times are all
distinct: see~(\ref{DwelPl})-(\ref{DwelTot}).

Then, if one considers the average velocity defined by
\begin{equation}
   V=d(\pi_+-\pi_-)/\tau,\label{DefVelocity}
\end{equation}
one finds that the result is the same in both cases, namely,
\begin{equation}
   V=\check{V}=d(1-\Gamma_N)\!\!\left/\mbox{$\sum_{n=0}^{N-1}$}r_n\right.,\label{VelNew}
\end{equation}
as
previously~\cite{fisher99,kolomeisky03,fisher05,fisher99a,kolomeisky00}.
This is not really surprising, because the asymptotic ratio of the
total distance to the total time should not depend on the way one
takes into account (or ignores) substeps. Note that $\Gamma_N=1$ is
the condition for stall, when $\pi_+=\pi_-$, in both analyses.

Now let us take a closer look at the ratio of splitting fractions,
because this is a quantity which one can readily obtain from
stepping observations to test theory. The effect of an external
force, $\vec{F}=(F_x,F_y,F_z)$, on the rates can be expressed in
leading order~\cite{fisher99,kim05} as
\begin{eqnarray}
&&u_l=u^0_l\exp(+d\mbox{\boldmath $\theta$}^{+}_{l}\mbox{\boldmath $\cdot$}\,{\bf F}/k_BT),\label{UexpF}\\
&&w_l=w^0_l\exp(-d\mbox{\boldmath $\theta$}^{-}_{l}\mbox{\boldmath
$\cdot$}\,{\bf F}/k_BT),\label{WexpF}
\end{eqnarray}
where $u^0_l,w^0_l$ are the load-free rates while the
load-distribution vectors $\mbox{\boldmath $\theta$}^+_l$ and
$\mbox{\boldmath $\theta$}^-_l$ serve to specify how the $2N$
distinct rates respond to the stress. The periodicity of the
stepping along the track (which we suppose is in the $x$ direction)
implies
\begin{equation}
\sum_{n=0}^{N-1}\left(\mbox{\boldmath
$\theta$}^{+}_{n}+\mbox{\boldmath $\theta$}^{-}_{n}\right)=\hat{\bf
e}_x=(1,0,0).\label{ThetaPerCond}
\end{equation}
Rearranging these relations leads to
\begin{eqnarray}
\Gamma_N&=&\prod_{n=0}^{N-1}\frac{w_n}{u_n}=\exp\left(-\frac{dF_x}{k_BT}\right)\prod_{n=0}^{N-1}\frac{w^0_n}{u^0_n}\nonumber\\
&=&\exp[-d(F_x-F_{S})/k_BT],\label{GammaFstall}
\end{eqnarray}
where the stall force is given by~\cite{fisher99,kim05}
\begin{equation}
F_{S}=\frac{k_BT}{d}\ln\left(\prod_{n=0}^{N-1}\frac{w^0_n}{u^0_n}\right).\label{DefFstall}
\end{equation}
From (\ref{SplOld}) we thus conclude
\begin{equation}
\ln(\check{\pi}_+/\check{\pi}_-)=-\ln\Gamma_N=d(F_x-F_{S})/k_BT.\label{LogGamma}
\end{equation}

This is a strong prediction since it asserts that the logarithm of
the stepping ratio $(\check{\pi}_+/\check{\pi}_-)$ depends linearly
on $F_x$ with a slope, $d/k_BT$, determined solely by the step size
$d$. However, this result depends crucially on the assumption that
the forward and back steps identified correspond to full enzymatic
cycles (as the haceks indicate). In a typical experiment, however,
 the hidden substeps result in a violation of
this assumption as we have explained.

Accordingly, let us, instead, compute the stepping fraction
$(\pi_+/\pi_-)$ for which~(\ref{StepSplProbPl})
and~(\ref{StepSplProbMi}) yield
\begin{equation}
\ln(\pi_+/\pi_-)=d^{*}(F_x-F_S)/k_BT\label{LogStepSplProb},
\end{equation}
where, for the sake of comparison with~(\ref{LogGamma}), we have
introduced an {\it effective step size}
\begin{equation}
d^{*}=\frac{-k_BT}{F_x-F_{S}}\ln\left(\frac{\Gamma_N+\Delta_{M,N}}{1+\Delta_{M,N}}\right).\label{DefdEffect}
\end{equation}
In general, this clearly depends on all the rates $\{u_j,w_j\}$ and
hence on the force $F_x$. In the special case $\Delta_{M,N}=0$,
however, $d^*$ reduces [via~(\ref{LogGamma})] to $d$. This condition
is, in fact, realized when $N=1$ since $\Delta_{M,1}$ vanishes
identically. But an $N=1$ model is unlikely to be adequate. Thus in
real systems neither the full linearity vs. $F_x$ nor the equality
$d^*=d$ are to be expected. In the vicinity of $F_S$, however, we
can estimate $d^*$ by expanding in powers of $\delta F_x\equiv
F_x-F_S$. This yields
\begin{eqnarray}
&&\Gamma_N=1-d\,\delta F_x/k_BT+O(\delta F_x^2),\label{GammaExpand}\\
&&\Delta_{M,N}=\Delta_S+\Delta'_S\delta F_x+O(\delta
F_x^2),\label{DeltaExpand}
\end{eqnarray}
where $\Delta_S=\Delta_{M,N}(F_x\!\!\,=\!F_S)$ while $\Delta'_S$ is
the corresponding derivative. Thence we find
\begin{equation}
d^{*}=\frac{d}{1+\Delta_S}+O(\delta F_x).\label{dEffectExpand}
\end{equation}
It follows from the definition~(\ref{DefDelta}) that $\Delta_S$
cannot be negative so that, quite generally, one has $d^*\leqslant
d$.

On the other hand, to be concrete, consider an ($N=2$, $M=1$) model
for which we have
\begin{align}
\!\!\Delta_S&=(w_0/u_0)_S=(u_1/w_1)_S\nonumber\\
&=(u^0_1/w^0_1)\exp\{d[F_{S}-(\mbox{\boldmath
$\theta$}^{+}_0+\mbox{\boldmath $\theta$}^{-}_0)\,\mbox{\boldmath
$\cdot$}\,{\bf F}_S]/k_BT\},\label{dDelta0forN2}
\end{align}
where the subscript $S$ denotes evaluation at ${\bf
F=F}_S=(F_x\!\!=\!F_S,F_y,F_z)$ in which $F_y$ and $F_z$ need not
vanish at stall~\cite{kim05}. It is evident, that $\Delta_S$ is not
bounded above so that $d^*/d$ is not bounded below. Indeed, the
experiments on kinesin of Nishiyama {\it et al.}~\cite{nishiyama03}
and of Carter and Cross on kinesin~\cite{carter05} lead to the
estimates $d^{*}\simeq 3.2\;\mathrm{nm}$ and $d^{*}\simeq
4.0\;\mathrm{nm}$, respectively, whereas the step size is $d\simeq
8.2\;\mathrm{nm}$. This clearly indicates the importance of the
hidden substeps in understanding the stepping fractions near stall.

As another concrete example we quote
\begin{eqnarray}
\Delta_S&=&\left(\frac{w_0}{u_0}+\frac{w_0w_1}{u_0u_1}+\frac{w_0w_1w_2}{u_0u_1u_2}\right)_S\nonumber\\
&=&\left(\frac{u_1u_2u_3}{w_1w_2w_3}+\frac{u_2u_3}{w_2w_3}+\frac{u_3}{w_3}\right)_S,
\end{eqnarray}
for an ($N=4$, $M=3$) model. From this, however, one sees that $d^*$
will be close to $d$ if $(u_3/w_3)_S$ and/or $(w_0/u_0)_S$ are
sufficiently small.

\begin{figure*}
\includegraphics[angle=0, scale=0.70]{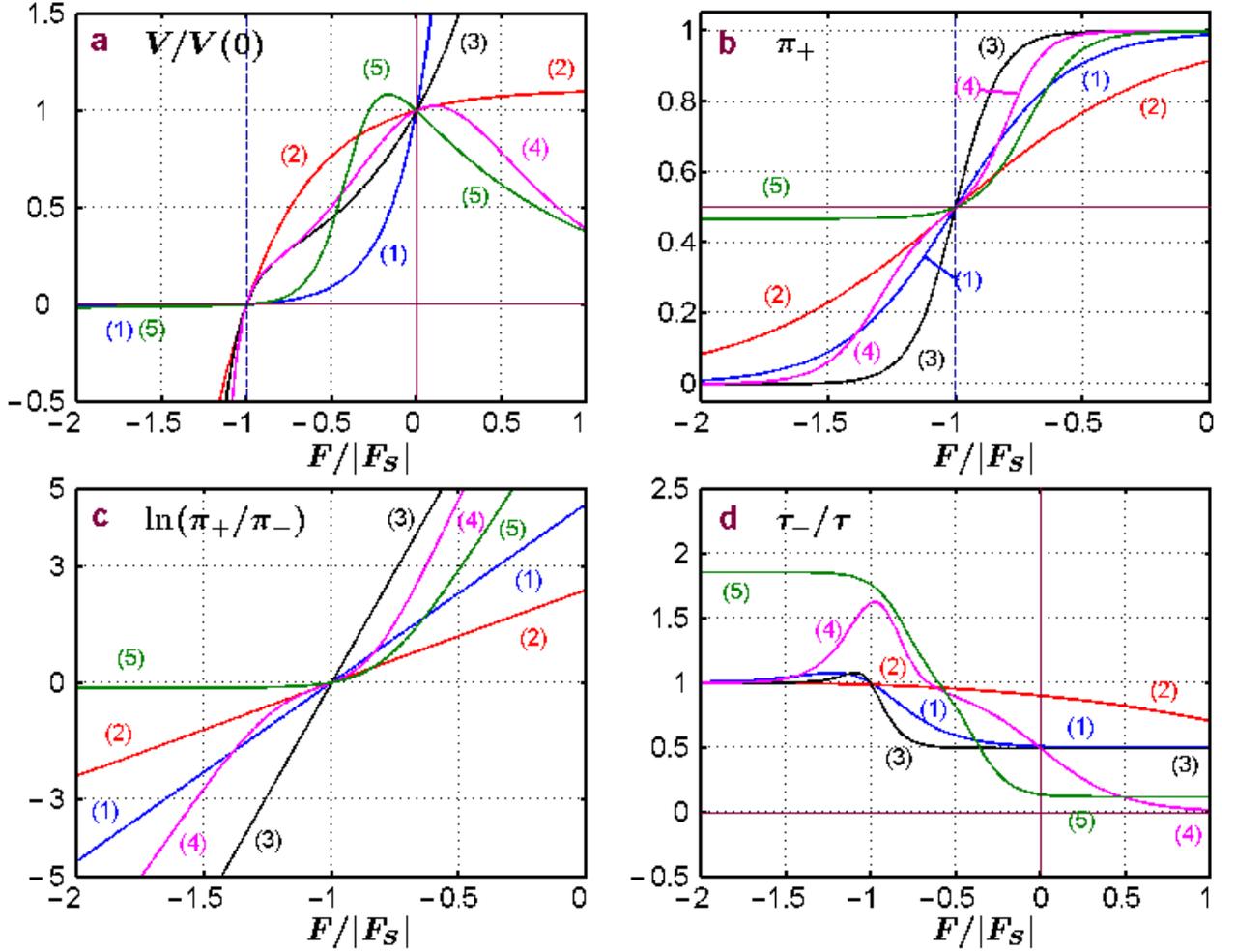}
\caption{\label{fig3:wide}Plots illustrating different dependencies
on the normalized force, $F/|F_S|$, of {\bf (a)} the normalized
velocity $V(F)/V(0)$; {\bf (b)} the forward splitting fraction
$\pi_+$; {\bf (c)} the logarithm of the splitting fraction ratio
$(\pi_+/\pi_-)$; and {\bf (d)} the ratio of mean prior backward to
mean total dwell time, $\tau_-(F)/\tau(F)$. The individual plots,
labeled (1) to (5), correspond to ($N=2$, $M=1$) models under scalar
loading, $\textbf{F}\!=\!(F,0,0)$, with the selected parameter
values presented in Table~I.  }
\end{figure*}

More generally it seems likely that both the full cycle assumption
and the hidden substep analysis should produce similar results when
the rates, $u_M$ and $w_{M+1}$, for the major transition are slow
relative to the substep rates. To demonstrate this explicitly let us
suppose that the latter rates satisfy
\begin{equation}
u_M=\eta\,\tilde{u}_M\quad\mbox{and}\quad w_{M+1}=\eta\,\tilde{w}_{M+1},\label{DefTildes}\\
\end{equation}
where $\eta$ is small while $\tilde{u}_M$ and $\tilde{w}_{M+1}$ and
all the substep rates, $u_l\equiv\tilde{u}_l$ $(l\neq M)$ and
$w_l\equiv\tilde{w}_l$ $(l\neq M+1)$, are held fixed. Then one finds
\begin{eqnarray}
\Gamma_{M,s}=\prod_{n=1}^{s}\frac{w_{M+n}}{u_{M+n}}&=&\eta\,\tilde{\Gamma}_{M,s}\quad\mbox{for}\quad
s\in[1,N-1],\nonumber\\
&=&\tilde{\Gamma}_N=\Gamma_N\quad\mbox{for}\quad s=N,\label{TildeGs}
\end{eqnarray}
where
$\tilde{\Gamma}_{M,s}=\prod_{n=1}^{s}(\tilde{w}_{M+n}/\tilde{u}_{M+n})$,
and similarly, via~(\ref{DefDelta}),
\begin{equation}
\Delta_{M,N}=\eta\sum_{p=1}^{N-1}\tilde{\Gamma}_{M,p}\equiv\eta\tilde{\Delta}_{M,N}.\label{TildeDelta}
\end{equation}
With the aid of these expressions we can rewrite the previous
results~(\ref{StepSplProbPl},\ref{StepSplProbMi})
and~(\ref{DwelTot}) as
\begin{eqnarray}
&&\pi_+=\frac{1+\eta\tilde{\Delta}_{M,N}}{1+\tilde{\Gamma}_N+2\eta\tilde{\Delta}_{M,N}}=1-\pi_-,\\
&&\tau=\check{\tau}\frac{1+\tilde{\Gamma}_N}{1+\tilde{\Gamma}_N+2\eta\tilde{\Delta}_{M,N}}.
\end{eqnarray}
Evidently to zero order in $\eta$, the analyses are equivalent as
anticipated.

As seen in earlier investigations that were confined to the velocity
vs. force relation~\cite{fisher99,fisher99a}, a surprisingly wide
range of behavior under varying loads is displayed even by the basic
$(N=2)$-state models when the rates are subject to the exponential
force distribution laws embodied
in~(\ref{UexpF}-\ref{ThetaPerCond})~\cite{fisher99,kim05}. This is
illustrated in Fig.3(a) which depicts the velocity $V$, normalized
by its value $V(0)$ under zero load, as a function of the imposed
load $F$ (supposed parallel to the $x$-axis) normalized by the stall
force magnitude $|F_S|$. The labeled plots (1) to (5) in
Fig.~\ref{fig3:wide} correspond to the selected parameter values
listed in Table~I. Note that {\it superstall loads} are included
($F/|F_S|<-1$) which for the parameter sets (1) and (5) results in
only a relatively small negative velocity (as uncovered in the
kinesin experiments of Carter and Cross~\cite{carter05}). Similarly,
{\it assisting loads} ($F/|F_S|>0$) are also covered and for sets
(2), (4), and (5) result in a saturating or, even, {\it decreasing}
velocity under increasing load. In the {\it substall} resistively
loaded region ($0>F/|F_S|>-1$) convex, concave, inflected, and even
nonmonotonic [see parameter set (5)] behavior is realized.

For these different cases the corresponding forward stepping
fractions $\pi_+(F)$ and the logarithmic ratios of forward/back
steps, ($\pi_+/\pi_-$), are depicted in Figs.~3(b) and (c). [Note
that in these figures only the {\it resisting} range of force,
$F<0$, is displayed.] Although the variation is always monotonically
increasing with $F$ (and $\pi_+=\pi_-=\frac{1}{2}$ when
$F\!=\!F_S$), a wide range of forms is evident. In the logarithmic
plot, Fig.3(c), one sees linear, concave, and inflected variation
close to stall. Furthermore, the value of the effective step size
$d^*$ varies markedly: see the last column in Table~I.

By using the explicit expressions in the Appendix the nature of
other statistical observables such as $\pi_{+-}(F)$, etc., is
readily explored. Experiments often measure the overall mean dwell
time, $\tau(F)$, between steps. Under assisting loads $(F>0)$, when
$\pi_-$ is negligible, $\tau(F)$ directly mirrors the reciprocal of
the velocity $V(F)$; but, in view of the factor $(\pi_+-\pi_-)$
in~(\ref{DefVelocity}), it varies somewhat differently under
resisting loads. More interesting is the behavior of the {\it
partial dwell time} $\tau_-(F)$ observed prior to back steps. This
is shown in Fig.~3(d) normalized by the overall dwell time
$\tau(F)=\pi_+\tau_++\pi_-\tau_-$ [see~(\ref{TotalDwellasSum})].
Even though, the ratio $\tau_-/\tau$ is confined to the range
$(0,2)$ beyond superstall (since $\tau_-/\tau<1/\pi_-$ and
$\pi_->\frac{1}{2}$ for $V<0$), striking nonmonotonic and inflected
variation arises.

Needless-to-say, many more plots exhibiting unexpected and
surprising behavior can be generated; but further exploration seems
most useful in connection with specific experimental data. Such
applications are planned.

\begin{table}
\caption{Parameter values for ($N=2$, $M=1$) models under scalar
loading, $F\equiv F_x$, employed in Fig.~\ref{fig3:wide} to
illustrate different force-dependencies. The last column presents
the computed values of the effective step size $d^*$.
[See~(\ref{DefdEffect}).]}
\begin{tabular}{|c|c|c|c|c|c|c|}
  \hline
  % after \\: \hline or \cline{col1-col2} \cline{col3-col4} ...
    & $d|F_S|/k_BT$ & $w^0_0/u^0_0$ & $w^0_1/u^0_0$ & $\theta^+_1$ & $\theta^-_0=\theta^-_1$ & $d^*/d$ \\\hline
  (1) & 9.2 & 10$^{-2}$ & 10$^{-2}$ & 0.5 & 0 & 0.501 \\
  (2) & 2.5 & 10$^{-2}$ & 10$^{-2}$ & 0 & 0.5 & 0.966 \\
  (3) & 23 & 10$^{-5}$ & 10$^{-5}$ & 0.07 & 0.43 & 0.503 \\
  (4) & 23 & 10$^{-5}$ & 10$^{-5}$ & $-$0.07 & 0.48 & 0.113 \\
  (5) & 10 & 3.4$\times$10$^{-4}$ & 2.5$\times$10$^{-3}$ & $-$0.1 & 0.1 & 0.118 \\
  \hline
\end{tabular}
\end{table}

%***********************************************************************************************************************************
\section{\label{sec:level5} Multiple observed transitions}

\begin{figure*}
\includegraphics[angle=0, scale=0.9]{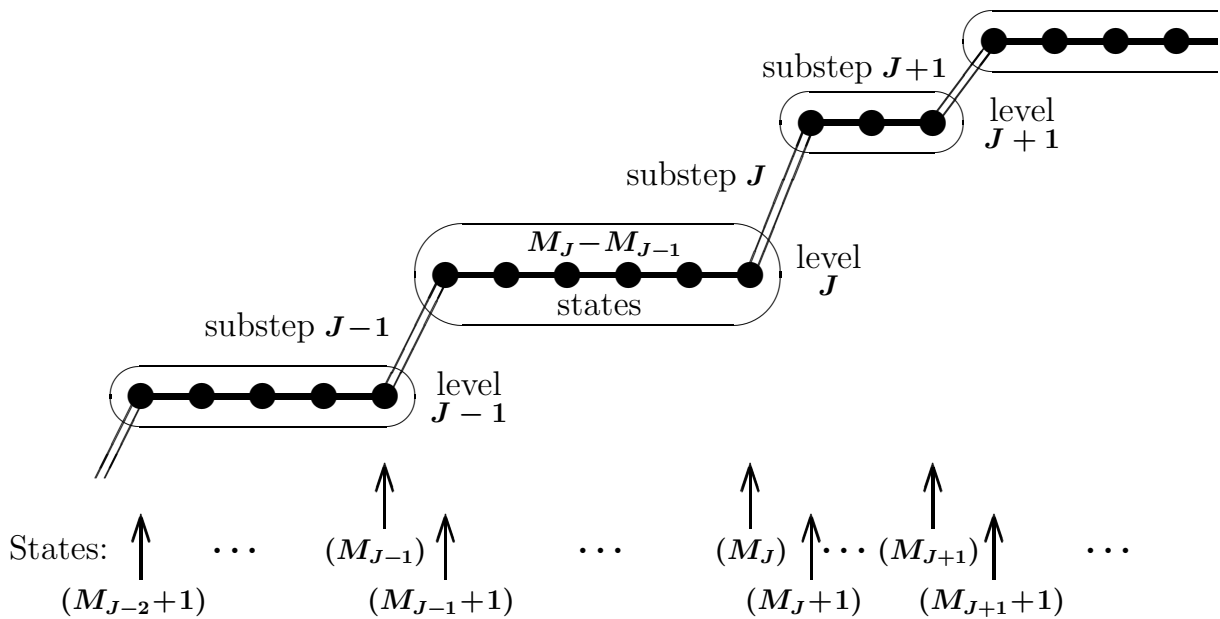}
\caption{\label{fig4:wide} Schematic depiction of individual
biochemical states, $(i)$, organized into mechanical levels, $A$,
$B$, $\cdots$, $J-1$, $J$, $\cdots$ when there are $K>1$ detectable
substeps or major transitions with substep $J$ going forward from
state $(M_J)$ to state $(M_J+1)$.}
\end{figure*}

In the previous sections we have derived expressions only for the
case of a {\it single} major transition in each enzymatic cycle;
that, indeed, is the typical situation for experiments on
conventional
kinesin~\cite{nishiyama02,block03,nishiyama03,carter05}. However,
our results can be generalized to the case in which several substeps
are sufficiently large to be clearly detected, while others remain
hidden in the noise. Suppose there are $K$ visible substeps of
(average) magnitudes $d_A$, $d_B$, $\cdots$, together totalling
\begin{equation}
d_A+d_B+\dots+d_K\simeq d,\label{SumVisibleStep}
\end{equation}
that occur between states ($M_J$) and ($M_{J}+1$) with, in sequence,
\begin{equation}
0\leq M_A<M_B<\dots<M_K\leq N-1.\label{SequenceOfSteps}
\end{equation}
Then between states ($M_{J-1}+1$) and ($M_{J}$) there are
$M_{J}-M_{J-1}$ hidden states and, consequently, all these states
belong (within the noise) to what we may call the same {\it
mechanical level}, $J$: see Fig.~\ref{fig4:wide}.

Now one can count all forward and backward detectable transitions in
a long run. Accordingly, let $n^J_{++}$ be the number of pairs of
observed substeps that enter the mechanical level $J$ via a $+$
transition, i.e., a step ($J-1$), and leave via a $+$ transition
taking a step $J$, and similarly for $n^J_{+-}$, $n^J_{-+}$, and
$n^J_{--}$. Then, as previously, we can estimate pairwise splitting
fractions for level $J$ via
\begin{eqnarray}
\pi^J_{++}&=&\frac{n^J_{++}}{n^J_{++}+n^J_{+-}}=1-\pi^J_{+-},\\
\pi^J_{-+}&=&\frac{n^J_{-+}}{n^J_{-+}+n^J_{--}}=1-\pi^J_{--}.
\end{eqnarray}
Likewise, we can introduce pair-wise dwell times,
$\tau^J_{\rho\sigma}$, for each mechanical level $J$ as the mean
times spent between a $\rho$ and $\sigma$ substep into and out of
that level.

With these definitions we may adopt the same approach used in
Sec.~\ref{sec:level3} by setting absorbing boundaries at states
$(L)=(M_{J-1})$ and $(R)=(M_J+1)$ around each level $J$ and studying
the appropriate first-passage processes. If we set
%\begin{eqnarray}
\begin{align}
N_J=\;&|M_A\!-\!M_B|,\;|M_B\!-\!M_C|,\;\cdots,\nonumber\\
&|M_K\!-\!M_{K-1}|,\;|N\!-\!M_K\!+\!M_A|,\nonumber\\
&\mbox{for}\quad J=A,\;B,\;\cdots,\;K-1,\;K,
\end{align}
%\end{eqnarray}
we can then conclude, using the previous results
(\ref{SplProbPlPl},\ref{SplProbMiPl}), that
\begin{equation}
\pi^J_{\rho\sigma}=\pi_{\rho\sigma}(M_{J-1},N_{J-1}).
\end{equation}
Similarly, recalling the definitions
(\ref{SolutionQPl},\ref{SolutionQMi}) for $T^+_j(M,N)$ and
$T^-_j(M,N)$ and the results (\ref{DwelPlPl}-\ref{DwelMiMi}), we
obtain
\begin{equation}
\tau^J_{\rho\sigma}=\tau_{\rho\sigma}(M_{J-1},N_{J-1}).
\end{equation}

The counting of individual hidden substeps developed in
Sec.~\ref{sec:level3e} can be carried forward to obtain the mean
number of substep transitions between two specified major
transitions. The results (\ref{nDDeven}-\ref{nRSodd}) essentially
apply directly with $M\Rightarrow M_{J-1}$ and $N\Rightarrow
N_{J-1}$.

In terms of the {\it conditional individual state dwell times}
introduced in (\ref{DefTPlPl}-\ref{DefTPlMi}) we also have
\begin{equation}
\tau^J_{\rho\sigma}=\sum_{k=M_{J-1}+1}^{M_J}T_k^{\rho\sigma}(M_{J-1},N_{J-1}).
\end{equation}
as a measure of the observable mean overall time spent in the
mechanical level $J$ subject to the ($\rho,\sigma$) conditions.

As regards potential applications of these results, the case $K=2$
may be reasonable for a first analysis of data for myosin V where,
as mentioned, a significant observable substep was originally
predicted~\cite{kolomeisky03} and later observed~\cite{uemura04};
however, the experiments also suggest~\cite{uemura04,baker04} that
stepping may proceed through two (or more) alternative pathways so
that a purely sequential model (to which our attention has been
restricted) may be inadequate~\cite{foot0A}. For the F$_1$-ATPase
motor~\cite{nishizaka04,shimabukuro03,ueno05} substeps have also
been reported and occasional back steps have been observed. Thus our
results should be applicable.

%***********************************************************************************************************************************
\section{\label{sec:level6}Summary and conclusion}

As explained in the Introduction, the need to develop the hidden
substep analysis we have presented arises from the fact that
experimentally detectable steps in the motion of a motor protein
along its track do not necessarily delineate the completion of full
biochemical enzymatic cycles. As a consequence, previous analyses
that addressed such observable statistics as back-stepping
fractions, $\pi_-$, and mean dwell times, $\tau_+$ and $\tau_-$,
measured prior to forward (or $+$) and back (or $-$) steps, were not
adequate to relate the underlying rates in a biomechanochemical
model, say $u_i$ and $w_j$ to the experimental data.

We have considered the basic $N$-state sequential kinetic model set
out in~(\ref{BasicModel}) and specified by $N$ forward rates $u_i$
from biochemical state $(i)$ and $N$ reverse rates $w_j$ from state
$(j)$. In general, the rates depend on the concentration of various
reagents [see, e.g.,~(\ref{ATPdepend})] and, in particular, vary
experimentally with the load force ${\bf F}=(F_x,F_y,F_z)$:
see~(\ref{UexpF},\ref{WexpF}).

The basic problem may then be set up by supposing that as the motor
progresses (or retrogresses) along its molecular track only a single
``major transition'' from state $(M)$ to $(M+1)$ $\:(0\leq M<N)$, or
its reverse, corresponds to a ``visible'' or detectable ``step'' in
the $N$-state cycle. All the other transitions are ``hidden'': see
Fig.~\ref{fig:wide}. [Cases in which more than one major transition
or observable (sub)step occur in each full enzymatic cycle are
analyzed in Sec.~\ref{sec:level5}.] It then transpires that two
crucial combinations of the rates $\{u_i,w_j\}$  play a central
role, namely, $\Gamma_N$ and $\Delta_{M,N}$ as defined in
(\ref{DefGamma})-(\ref{DefDelta}).

Indeed, explicit expressions for the forward and backward stepping
fractions, $\pi_+$ and $\pi_-$, are derived in terms of $\Gamma_N$
and $\Delta_{M,N}$ in Sec.~\ref{sec:level3b} and presented
in~(\ref{StepSplProbPl}) and (\ref{StepSplProbMi}). It proves
helpful, furthermore, to relate $\pi_+$ and $\pi_-$ to the {\it
conditional} or {\it pairwise step probabilities}, $\pi_{++}$,
$\pi_{+-}$, etc., for $+$ steps followed by a $+$ step, or by a $-$
step, etc., that can be defined via counting observations, as
explained in~(\ref{PairwiseSplNorm})-(\ref{StepSplInPairSplMi}) and
the associated text. These pairwise probabilities are likewise
expressible in terms of $\Gamma_N$ and $\Delta_{M,N}$:
see~(\ref{SplProbPlPl}) and (\ref{SplProbMiPl}).

From these results one can see -- as explained in further detail in
Sec.~\ref{sec:level4} -- that only when
$\;\Delta_{M,N}\ll\min\{1,\Gamma_N\}\;$ can one neglect the hidden
substeps without risk of serious error. Particularly instructive is
the variation of $\;\ln(\pi_+/\pi_-)\;$ as the load $F_x$ passes
through stall (at which $\;\pi_+=\pi_-=\frac{1}{2}\;$ so that the
mean velocity vanishes). One may then define an {\it effective step
length}, $d^*$, via~(\ref{LogStepSplProb}) and (\ref{DefdEffect}).
When $\Delta_{M,N}\rightarrow 0$ (or if hidden substeps are ignored)
one has the simple {\it equality} $\;d^*=d$, where $d$ is the full
step size of the motor (per cycle). But, in fact, $d^*/d$ must
always be {\it less} than unity and, as seen in experiment and
illustrated in Table~I, this ratio is typically of magnitude only
0.3 to 0.5.

Going beyond simple counts of backward and forward steps, one may
also define {\it conditional mean dwell times}, $\tau_{++}$,
$\tau_{+-}$, etc., for the time spent between a pair of successive
$+$ steps, or a $+$ step followed by a $-$ step, etc.:
see~(\ref{DefPairwiseDwells}). These pairwise mean times can
likewise be calculated [see~(\ref{DwelPlPl}) to (\ref{DwelMiMi})] in
terms of {\it individual-state post dwell times}, $T_j^+(\equiv
T_j^{+\diamond})$ and $T_j^-(\equiv T_j^{-\diamond})$, that
represent the mean time spent in a state $(j)$ following a $+$ or
$-$ step, respectively: see Sec.~\ref{sec:level3d} for a fuller
explanation of the notation, etc. The corresponding explicit
expressions, (\ref{SolutionQPl}) and (\ref{SolutionQMi}), involve
the basic rate products $\Gamma_{l,m}$ and their sums,
$\Delta_{l,n}$, as again defined in (\ref{DefGamma}) and
(\ref{DefDelta}). The final results for $\tau_+$, $\tau_-$, and for
the overall mean dwell time $\tau$ (between $+$ or $-$ steps) for
general $M$ and $N$ entail slightly simpler sums:
see~(\ref{ReDefQ})-(\ref{DwelTot}).

More transparent formulae for the stepping fractions and dwell times
for $N=2$ models (involving only the rates $u_0$, $u_1$, $w_0$, and
$w_1$,) and for selected $N=4$ models, are presented in Appendix~A.
In addition, the parts of Fig.~\ref{fig3:wide} and the associated
discussion in Sec.~\ref{sec:level4}, illustrate that a wide range of
different types of behavior of $\pi_+(F)$, $\ln[\pi_+(F)/\pi_-(F)]$,
and $\tau_-(F)$ as functions of load can be realized even within
simple $N=2$ models.

At a higher level of detail, conditional individual-state dwell
times, $T_j^{++}$, $T_j^{+-}$, $\cdots$, can be derived
[see~(\ref{DefTPlPl})-(\ref{DefTPlMi})] and, likewise, {\it post}
(as against the previously mentioned {\it prior}) dwell times,
$\tau_{+\diamond}$ and $\tau_{-\diamond}$:
see~(\ref{PostDwelAsSumT}). Finally, one can obtain the
expressions~(\ref{nDDeven}), (\ref{nDDodd}), (\ref{nRSeven}), and
(\ref{nRSodd}), for the mean number, $n_{j,+}^{\rho\sigma}$ and
$n_{j,-}^{\rho\sigma}$ of hidden, forwards and backwards, substeps
from an individual state $(j)$ that occur in a time interval between
detectable steps, i.e., major transitions specified by
$\{\rho,\sigma\}=\{\cdots,+-,-+,\cdots\}$:
see~(\ref{9combinations}). These results provide quantitative
estimates for the number of ``lost'' or ``hidden'' transitions
occurring in an enzymatic cycle. Such information could be of
particular interest for real motor proteins since, when operating in
cells to achieve mitosis, etc., they may frequently be in
close-to-stall conditions where reverse substeps are likely to be
most frequent~\cite{grill03,pecreaux06}.

In conclusion, we have provided a detailed analysis of the
statistics of mechanochemical transitions that must be hidden in the
experimental noise when a motor protein on its track moves
processively via distinct steps, or reaches stall. As experimental
resolution at the microsecond and nanometer scales improves, we can
expect that such analyses will be increasingly valuable for
extracting reliable inferences about motor mechanisms from
observational data.

\begin{acknowledgments}
The authors much appreciate correspondence and interactions with
Dr.~R.~A.~Cross and Dr.~N.J.~Carter. Support from the National
Science Foundation under Grant No. CHE 03-01101 is acknowledged.
M.~L. is grateful for the financial support of the Royal Institute
of Technology and the Wallenberg Foundation and for the hospitality
of the Institute for Physical Science and Technology in Fall 2005.
\end{acknowledgments}

%***********************************************************************************************************************************
\appendix
\section{\qquad Expressions for Two-State and Four-State Models}

For convenience of reference we provide here explicit expressions
for $N=2$ models with $M=1$. First, we recall the full-cycle
expressions~\cite{kolomeisky03,kolomeisky05}
\begin{equation}\label{PiFC}
    \check{\pi}_+=1-\check{\pi}_-=\frac{u_0u_1}{u_0u_1+w_0w_1},
\end{equation}
\begin{equation}\label{TauFC}
    \check{\tau}_+=\check{\tau}_-=\check{\tau}=\frac{u_0+u_1+w_0+w_1}{u_0u_1+w_0w_1},
\end{equation}
the result for general $N$ being given
in~(\ref{SplOld}),(\ref{DwelOld}). Allowing for hidden substeps
leads to
\begin{equation}\label{PiN2}
    \pi_+=1-\pi_-=\frac{u_1(u_0+w_0)}{u_0u_1+w_0w_1+2u_1w_0},
\end{equation}
while the distinct prior dwell (or stepping) times are given by
\begin{equation}\label{TauPlN2}
    \tau_+\equiv\tau_{\diamond+}=\frac{(u_0+w_0)^2+u_0(u_1+w_1)}{(u_0+w_0)(u_0u_1+w_0w_1+u_1w_0)},
\end{equation}
\begin{equation}\label{TauMiN2}
    \tau_-\equiv\tau_{\diamond-}=\frac{(u_1+w_1)^2+w_1(u_0+w_0)}{(u_1+w_1)(u_0u_1+w_0w_1+u_1w_0)},
\end{equation}
with the mean overall dwell time
\begin{equation}\label{TauN2}
    \tau=\pi_+\tau_++\pi_-\tau_-=\frac{u_0+u_1+w_0+w_1}{u_0u_1+w_0w_1+2u_1w_0}.
\end{equation}
For general $N$ see~(\ref{StepSplProbPl}),~(\ref{StepSplProbMi}),
and~(\ref{DwelPl})-(\ref{DwelTot}).
%---------------------------------------------------------------

The partial or conditional pairwise step probabilities follow, by
specializing~(\ref{SplProbPlPl}) and~(\ref{SplProbMiPl}), as
\begin{equation}\label{PiPlN2}
    \pi_{++}=1-\pi_{+-}=\frac{u_0u_1}{u_0u_1+w_0w_1+u_1w_0},
\end{equation}
\begin{equation}\label{PiMiN2}
    \pi_{-+}=1-\pi_{--}=\frac{u_1(u_0+w_0)}{u_0u_1+w_0w_1+u_1w_0}.
\end{equation}
The denominators here, say
\begin{equation}\label{DenomD2}
    \mathfrak{D}_2=u_0u_1+w_0w_1+u_1w_0,
\end{equation}
should be contrasted with those in~(\ref{PiN2}) and (\ref{TauN2}),
namely,
\begin{equation}\label{DenomD2Pl}
    \mathfrak{D}^+_2=\mathfrak{D}_2+u_1w_0=u_0u_1+w_0w_1+2u_1w_0.
\end{equation}

At the next level of {\it individual} (or {\it partial}) (sub)state
properties, the individual or substate dwell times follow
from~(\ref{ReDefQ}), which yields
\begin{equation}\label{St01DwN2}
    T_0=(u_1+w_1)/\mathfrak{D}^+_2,\qquad T_1=(u_0+w_0)/\mathfrak{D}^+_2,
\end{equation}
which, in accord with~(\ref{DwelTot}), satisfy $\tau=T_0+T_1$. The
partial substate dwell times, given generally in
(\ref{SolutionQPl},\ref{SolutionQMi}), are
\begin{equation}\label{St01PlDwN2}
    T_0^+=(u_1+w_1)/\mathfrak{D}_2,\mathfrak{\qquad} T_1^+=u_0/\mathfrak{D}_2,
\end{equation}
\begin{equation}\label{St01MiDwN2}
    T_0^-=w_1/\mathfrak{D}_2,\mathfrak{\qquad} T_1^-=(u_0+w_0)/\mathfrak{D}_2.
\end{equation}
%---------------------------------------------------------------
Finally, the conditional or pairwise mean dwell or stepping times
are
\begin{eqnarray}
    \tau_{++}&=&\tau_{--}=(u_0+u_1+w_0+w_1)/\mathfrak{D}_2,\label{TauPlPlN2}\\
    \tau_{+-}&=&[(u_1+w_1)^2+u_0w_1]/(u_1+w_1)\mathfrak{D}_2,\label{TauPlMiN2}\\
    \tau_{-+}&=&[(u_0+w_0)^2+u_0w_1]/(u_0+w_0)\mathfrak{D}_2.\label{TauMiPlN2}
\end{eqnarray}

The mean numbers of {\it hidden transitions} follow from the results
in Sec.~\ref{sec:level3e}. For the simplest $N=2$ ($M=1$) case we
obtain
\begin{align}
    n_{0,+}^{++}-1&=n_{0,+}^{+-}=n_{0,+}^{-+}=n_{0,+}^{--}=n_{0,+}^{-\diamond}\nonumber\\
    &=n_{0,+}^{\diamond-}=n_{1,-}^{++}=n_{1,-}^{+-}=n_{1,-}^{-+}\nonumber\\
    &=n_{1,-}^{--}-1=n_{1,-}^{+\diamond}=n_{1,-}^{\diamond+}=u_0w_1/\mathfrak{D}_2,\\
    n_{0,+}^{+\diamond}&=n_{0,+}^{\diamond+}=u_0(u_1+w_1)/\mathfrak{D}_2,\\
    n_{1,-}^{-\diamond}&=n_{1,-}^{\diamond-}=w_1(u_0+w_0)/\mathfrak{D}_2,\\
    n_{0,+}^{\diamond\diamond}&=u_0(u_1+w_1)/\mathfrak{D}^+_2,\\
    n_{1,-}^{\diamond\diamond}&=w_1(u_0+w_0)/\mathfrak{D}^+_2.
\end{align}

To provide further insight we quote some results for $N=4$ models
with $M=3$, i.e., with the major step as the last transition from
state (3) to $[4\equiv0]$. Thus we have
\begin{equation}\label{PiN4}
    \pi_-=w_0\mathfrak{N}_0/\mathfrak{D}^+_4, \qquad \pi_+=u_3\mathfrak{N}_3/\mathfrak{D}^+_4,
\end{equation}
\begin{equation}\label{TauN4}
    \tau=(\mathfrak{N}_0+\mathfrak{N}_1+\mathfrak{N}_2+\mathfrak{N}_3)/\mathfrak{D}^+_4,
\end{equation}
where, to write the numerator and denominator contributions
compactly, we introduce the short-hand product notation
\begin{eqnarray}
    && u_{0,1}=u_0u_1,\; u_{0,1,2}=u_0u_1u_2,\;\cdots,\nonumber\\
    && w_{0,1}=w_0w_1,\; w_{0,1,2}=w_0w_1w_2,\;\cdots,\label{ShortHand}
\end{eqnarray}
etc. Then we have
\begin{eqnarray}
    \mathfrak{N}_0&=&u_{1,2,3}+w_1u_{2,3}+w_{1,2}u_3+w_{1,2,3},\\
    \mathfrak{N}_1&=&u_{2,3,0}+w_2u_{3,0}+w_{2,3}u_0+w_{2,3,0},\\
    \mathfrak{N}_2&=&u_{3,0,1}+w_3u_{0,1}+w_{3,0}u_1+w_{3,0,1},\\
    \mathfrak{N}_3&=&u_{0,1,2}+w_0u_{1,2}+w_{0,1}u_2+w_{0,1,2},\label{NumTauN4}
\end{eqnarray}
while the denominator is given by
\begin{eqnarray}
    \mathfrak{D}^+_4&=&w_0\mathfrak{N}_0+u_3\mathfrak{N}_3=u_{0,1,2,3}+2w_0u_{1,2,3}\nonumber\\
                &+&2w_{0,1}u_{2,3}+2w_{0,1,2}u_3+w_{0,1,2,3}.\label{DetN4}
\end{eqnarray}

For the purpose of comparison we quote the result for the velocity,
namely,
\begin{equation}\label{VelocityN4}
    V=d(u_{0,1,2,3}-w_{0,1,2,3})/(\mathfrak{N}_0+\mathfrak{N}_1+\mathfrak{N}_2+\mathfrak{N}_3),
\end{equation}
which, of course, is {\it not} simply $d/\tau$:
see~(\ref{DefVelocity}). Finally, then we also quote
\begin{align}
    \tau_{++}=\tau_{--}&=[\mathfrak{N}_0+\mathfrak{N}_1+\mathfrak{N}_2+\mathfrak{N}_3\nonumber\\
                       &+w_0u_3(u_1+w_1+u_2+w_2)]/\mathfrak{D}_4,\label{TauPlPlN4}
\end{align}
and
\begin{equation}\label{TauPlN4}
    \tau_+=\frac{u_{2,1,0}\mathfrak{N}_0+u_{2,1}\mathfrak{D}_1\mathfrak{N}_1+u_2\mathfrak{D}_2\mathfrak{N}_2+\mathfrak{D}_3\mathfrak{N}_3}{\mathfrak{D}_4\mathfrak{N}_3},
\end{equation}
where
\begin{align}
    \mathfrak{D}_N&=u_{0,1,\cdots,N-1}+w_0u_{1,2,\cdots,N-1}\nonumber\\
                &+w_{0,1}u_{2,3,\cdots,N-1}+\cdots+w_{0,1,\cdots,N-1},\label{DetNGen}
\end{align}
from which $\tau_-$ follows by using $\pi_-\tau_-+\pi_+\tau_+=\tau$.
Of course, results for $M=0$, $1$, and $2$ and for $\pi_{++}$,
$\tau_{++}$, $T_j$, etc., follow from the expressions derived in
Sec.~\ref{sec:level3}.
%----------------------------------------------------------------
%\bibliography{chameleonIII}
%\bibliographystyle{unsrt}
%\bibliographystyle{plain}

\end{document}